\documentclass[preprint,authoryear,12pt]{elsarticle}

\usepackage{mathtools}
\usepackage{amsmath}
\usepackage{amssymb}
\usepackage{ushort}% matrix symbols
\usepackage{esint}
\usepackage{psfrag}
\usepackage{trfsigns}
\usepackage{subfig}
\journal{Wave Motion}

\begin{document}

\begin{frontmatter}

%% Title, authors and addresses

%% use the tnoteref command within \title for footnotes;
%% use the tnotetext command for the associated footnote;
%% use the fnref command within \author or \address for footnotes;
%% use the fntext command for the associated footnote;
%% use the corref command within \author for corresponding author footnotes;
%% use the cortext command for the associated footnote;
%% use the ead command for the email address,
%% and the form \ead[url] for the home page:
%%
%% \title{Title\tnoteref{label1}}
%% \tnotetext[label1]{}
%% \author{Name\corref{cor1}\fnref{label2}}
%% \ead{email address}
%% \ead[url]{home page}
%% \fntext[label2]{}
%% \cortext[cor1]{}
%% \address{Address\fnref{label3}}
%% \fntext[label3]{}

\title{Predicting the statistics of wave transport through chaotic cavities by the Random Coupling Model: a review and recent progress}

%% use optional labels to link authors explicitly to addresses:
%% \author[label1,label2]{<author name>}
%% \address[label1]{<address>}
%% \address[label2]{<address>}

\author{Gabriele Gradoni, Jen-Hao Yeh, Bo Xiao, Thomas M. Antonsen,\\ Steven M. Anlage, and Edward Ott}

\address{Institute for Research in Electronics and Applied Physics,\\ 
 Department of Physics, \\
 Department of Electrical and Computer Engineering\\ University of Maryland,\\ College Park, MD-20742, USA}

\begin{abstract}
%% Text of abstract
In this review, a model (the Random Coupling Model) that gives a statistical description of the coupling of radiation into and out 
of large enclosures through localized and/or distributed channels is presented. 
The Random Coupling Model combines both deterministic and statistical phenomena. 
The model makes use of wave chaos theory to extend the classical modal description of the cavity fields in the presence of boundaries that lead to chaotic ray 
trajectories. 
The model is based on a clear separation between the universal statistical behavior of the isolated chaotic system, and the deterministic  
coupling channel characteristics.
Moreover, the ability of the random coupling model to describe interconnected cavities, aperture coupling, and the effects of  
short ray trajectories is discussed. 
A relation between the random coupling model and other formulations adopted in acoustics, optics, and statistical electromagnetics, is examined.
In particular, a rigorous analogy of the random coupling model with the Statistical Energy Analysis used in acoustics is presented.
\end{abstract}

\begin{keyword}
%% keywords here, in the form: keyword \sep keyword
Electromagnetic environment (EME), complex cavity, wave chaos, statistical modeling, 
impedance matrix, admittance matrix, short-range orbits, antenna, aperture, interconnected systems, 
statistical energy analysis (SEA).
%% MSC codes here, in the form: \MSC code \sep code
%% or \MSC[2008] code \sep code (2000 is the default)

\end{keyword}

\end{frontmatter}

% \linenumbers

%% main text
\section{Introduction}
\label{sec:introduction}
The random coupling model (RCM) addresses the problem of statistically modeling the wave behavior of large irregular cavities connected 
to an external environment by one or more channels. 
The RCM is formulated in terms of impedance, a concept common to both the electromagnetic and acoustics communities.  
As such, many of the lessons learned in an electromagnetic context presented below have close analogs in the field of vibro-acoustics.

Increasingly complex scenarios in electronics and telecommunications make the detail of structures and circuitry more and more difficult to model. 
In addition, often in optics, electronics, and acoustics, the need for higher bit-rates pushes 
wave sources to emit at very short wavelengths compared to the characteristics size of the excited system. 
In this regime, the scattering process can be very sensitive to details. 
A statistical approach then becomes appropriate. Specifically, one can ask what the statistics of quantities of interest are relative to a suitable 
random choice of the system. This is the goal of the RCM.
Other statistical approaches have already been used to study reverberation chambers and wireless propagation channels. 
For example, a Gaussian state is assumed \emph{a priori} for  
the complex Cartesian field upon maximization of entropy \cite{hill2009}. 

The RCM approach involves two physical prescriptions inherited from the 
quantum chaos theory of compound nuclei in nuclear reaction theory \cite{1967_Wigner,Mitchell01RMP_2010}. 
The first prescription is to replace the exact spectrum of the cavity with a spectrum whose statistics are obtained from those of a suitable ensemble 
of random matrices. 
The second prescription, motivated by the typical chaotic behavior of rays in complex enclosures, is to take for the mode amplitude a superposition of random plane-waves. 

The applicability of Random Matrix Theory (RMT) to the solution of simple wave equations in bounded regions where the corresponding ray trajectories are 
chaotic was originally explored by \cite{MacDonold1979}.

Understanding the influence of ports (antennas/sensors) and apertures on cavity field dynamics is itself interesting in practice. 
The chaotic regime of fields inside an irregular cavity is established by the boundary geometry driving ray propagation. 
From experimental observations, it is known that this results in a very high variability of the wave properties of a system.
The RCM incorporates this in a formally elegant and yet physically meaningful way.

The main goal in formulating the RCM was to connect the behavior predicted by random matrix theory (RMT) developed for the case of isolated wave chaotic systems, e. g., 
closed cavities, to the study of the behavior of such a system when it is connected to other systems through deterministic 
wave propagation channels, and to investigate the resulting 
input/output system characteristics. 
This was done in the quantum mechanical context, for example, by \cite{Brouwer1995}.
The RCM was developed for resonators with spatially localized channels \cite{Zheng2005T}. 
In 2006, studies were published for both the single channel \cite{Zheng2006a}, and the multiple channel \cite{Zheng2006b} cases. 
Those two papers developed previous studies considering isolated chaotic systems exhibiting Hamiltonian chaos \cite{Ott2002B}. 

The RCM attacks this problem from the point of view of port impedance/admittance matrices, in which the universal fluctuation of fields within the cavity is joined to the specific 
radiation characteristics of deterministic channels, whose geometry is presumed to be known \emph{a priori}.
 
%Irregular microwave cavities constitute a striking example of this class of systems, and they have been widely used to emulate the spectral behavior of complex 
%compound nuclei \cite{MacDonold1979,Bohigas1984,Stockmann1990}. 
%RCM is given in terms of channel impedance as a decomposition theorem where the presence of chaotic scattering introduces a ``universal'' fluctuation 
%affecting the radiation resistance of the channel, i.e., the real part of the terminal impedance when the metallic parts of the channel radiate in free-space \cite{Yeh2010a}.
\section{Mathematical model}
The wave chaos inside the cavity is expressed as a universal (i.e., system independent) superposition of  ``chaotic'' modes resulting in a universal ensemble of 
matrices, and the presence of channels connected with the external 
environment results in ``dressing'' this superposition with the radiation impedance matrix $Z^{rad}$ of channel terminals. 
In the case of a cavity with a single port the cavity impedance $Z^{cav}$ can be expressed as 
\begin{equation}\label{eqn:Zcav_3D_one}
 Z^{cav} = i \Im \left \{ Z^{rad} \right \} + \Re \left \{ Z^{rad} \right \} \, \xi \,\, ,
\end{equation}
where $Z^{\left ( rad \right )}$ is a port radiation impedance that will be defined later in the paper, $\xi$ is a statistically fluctuating variable corresponding to a sum over chaotic modes, 
and $\Im \left \{ \cdot \right \}$ and $\Re \left \{ \cdot \right \}$ denote imaginary and real parts. 
Thus, the statistics of $Z^{cav}$ is determined by the statistics of the random matrix $\xi$ and by the nonstatistical matrix $Z^{rad}$.
The sum over chaotic modes is universal in the sense that it depends only on a single parameter capturing cavity losses and modal structure through the average 
quality factor \cite{2003_Arnaut} and the mean spacing between nearest neighbor eigenmodes \cite{1967_Wigner} and on no other cavity details.

Recently, the RCM has been developed for three dimensional cavities with distributed ports, e.g., complicated antennas, in the same form of (\ref{eqn:Zcav_3D_one}) \cite{Antonsen_2011}. 
The presence of apertures calls for a different modeling attack, leading to the same form of the RCM for the cavity admittance matrix. 
The problem of external radiation coupling to the cavity through an aperture has been solved \cite{2012_GradoniA,2012_GradoniB}. 
The RCM introduces and incorporates system-specific characteristics efficiently.
It will be also shown how deviations from pure wave chaos can be included in the RCM by considering short-ray orbits \cite{Ishio1995,Baranger1996}. 
The RCM impedance expression (\ref{eqn:Zcav_3D_one}) is preserved and the non-universal term is augmented by tracing those ``non-chaotic'' short ray orbits \cite{Hart2009}.

In realistic physical scenarios one may employ different kinds of channel coupling to a closed electromagnetic environment (EME). 
Imagine a situation involving compartments of ships/aircraft/automobiles/trains: there could be present very localized terminals belonging to circuitry located inside bays, 
as well as very large windows and ports. 
It is thus quite useful to formulate a ``hybrid'' RCM involving both port terminals and electrically wide apertures, and to consider interconnected 
cavities. 

In this paper, the general formulation of the RCM for ports and apertures is reviewed.
Then, the predictions of probability distribution functions of desired input/output parameters (e.g., impedances, admittances or scattering parameters) obtained for single-mode channels 
by Monte Carlo computation of the RCM, and their experimental confirmation by measurements on actual systems are recalled. 
Finally, an application of the model to a situation of practical interest is discussed: the modeling of linear chains of interconnected chaotic cavities \cite{gradoni2012PRE}.
Interestingly, the formula obtained for the power flowing through a chain of lossy chaotic cavities is similar to the statistical energy analysis (SEA) 
approach developed for the propagation of acoustic and vibrational waves. 
\section{Historical remarks and methods}
\label{sec:methods}
Beginning in 1982 there was  rapid development in the application of RMT to both quantum and classical chaotic systems.
The first use of RMT was devoted to reproducing the spacing distribution of
energy levels in compound nuclei.  Later Rydberg levels of the hydrogen atom in a
strong magnetic field, and elastomechanical eigenfrequencies of irregularly
shaped quartz blocks were studied, as well as fluctuations of conduction through mesoscopic wires in a magnetic
field \cite{Guhr1998,Weiden_RMP_2009,Mitchell01RMP_2010}.
Studies on quantum chaos of quantum transport are referenced in the reviews \cite{Beenakker1997,2000_Alhassid}.
However, only a very small number of experiments on wave-chaotic scattering and
eigenfrequencies of electromagnetic cavities existed.
The first studies of irregularly shaped microwave cavities \cite{Stockmann1990,Doron1990,Sridhar1991,1992_graf} paved the way to electromagnetic wave-chaotic scattering
research. Microwave cavities with irregular shapes (having ray trajectories evolved 
by a chaotic map)
provided a simple and effective physical framework for the study of wave-chaos, where not only the
magnitude, but also the phase of scattering coefficients, can be directly measured
from experiments. 
The picture of rays is analogous to that of particles bouncing in a confined system with irregular potential barriers, whose trajectories exhibit exponential divergence in phase space, i.e., 
two particles originating from the same position with slightly different linear momentum follow very different paths after a few bounces off the boundaries. 
This dynamic effect has an impact on the spectrum of complex wave systems that are chaotic in the classical limit, which can be modeled by an unpredictable (random) Hamiltonian. 

The derivation of the RCM exploits both the Wigner surmise \cite{1967_Wigner} and the Berry hypothesis \cite{Berry1977}. 
If the dimensions of the cavity are much greater than the excitation wavelength, the semiclassical regime can be invoked.
Irregular boundaries create highly disordered mode topologies.
Therefore, the mode amplitude is locally modeled as an isotropic random superposition of plane waves (the Berry hypothesis). 
The motivation of this hypothesis is that the orbits of rays in a fully chaotic system visit all regions of phase space ergodically. 
In particular, if some time along a long orbit is randomly chosen, the probability density of the orbit direction is uniform in $0$ to $2 \pi$.
At the same time, the resonant behavior of modes is preserved by using RMT to generate mode wavenumbers (the Wigner surmise). 
Here, eigenvalues of a random matrix of the Gaussian Orthogonal Ensemble (GOE) with Gaussian distributed elements have been used since the time-reversal invariance is assumed to hold. 
%The associated eigenvectors have been termed ``chaotic modes''. 
The present approach is not phenomenological as the invoked ensemble when using RMT expresses the symmetry properties of the chaotic class a cavity belongs to 
(see \cite{So_1995} for experimental work where ``time reversibility'' is violated by the presence of a magnetized ferrite).

This motivates the use of the Gaussian ensembles.
However, in the presence of losses and interactions with the external environment, e.g., reactions in compound nuclei \cite{Mitchell01RMP_2010}, the universal properties of a system are intimately modified. 
For resonant electromagnetic systems, a unified model describing the behavior of open systems with losses was generated by several groups \cite{Savin2001,Fyodorov2004,Savin2004,Fyodorov2005,Savin2005}. 
Several extensions and experimental validation were published 
\cite{Hemmady2005a,Hemmady2005b,Hemmady2006a, Hemmady2006b,Zheng2006c,Yeh2010a,Yeh2010b,Yeh2012a,Yeh2012b,Hemmady2012} 
related to the theory of \cite{Zheng2006a,Zheng2006b,Zheng2006c}.  
After that, an effort was made to incorporate imperfections arising from the interaction of ports with nearby objects and cavity walls, giving rise to short-orbits \cite{Hart2009}. 
%A number of applications to scattering in superconducting cavities, and fading in communication systems \cite{Sirko_2004,Yeh2012a} were also presented to support experimental confirmation of the model.
More recently, the model has been extended to 3D cavities with distributed ports \cite{Antonsen_2011}, and to account for the interaction through arbitrary apertures \cite{Gradoni_EMCRoma_2012}.
Interestingly, the electrical formulation underlying the RCM is formally and physically similar to other formulations involving complex wave systems of different physical nature, e.g., acoustics. 
In the following, the analogy between the RCM and statistical approaches to understanding the properties of complex wave structures in applied electromagnetics, 
acoustics \& vibrations, and random lasers is briefly discussed.

% REVISE DEFINITION..PUT CLOSE TO STATISTICAL PHYSICS
In statistical electromagnetics (SEM), statistics and stochastic analysis are used to model complex electromagnetic 
systems \cite{Holland1999B}. The aim is, besides conceiving simple and effective design tools, to shed light on the collective behavior of modes and rays when subject to irregular 
and heterogeneous boundaries. 
Cavities and resonators constitute a natural and fertile terrain for SEM to be developed and applied. A number of models have been proposed to explain and predict 
fluctuations inside cavities (see for example \cite{hill2009}). The RCM improves on those phenomenological theories by describing 
random fluctuations through statistical ensemble theories \cite{schwabl2006}. In contrast, the RCM builds on the exact model of impedance and 
admittance matrices as derived from Maxwell's equations. 

As previously remarked, the use of RMT in physical systems dates back to 1950, when \cite{Wigner_Dirac_1950} adopted symmetry properties of statistical ensembles to model the Hamiltonian 
of complex nuclear systems. Besides nuclear theory, early applications of RMT can be found in acoustics and vibration systems \cite{Couchman_1992,2007_Tanner,Weaver_2010}.  
The pure statistical perspective of physical systems is even older and originates from thermodynamics. Interestingly, this perspective led spontaneously to the use of random matrices 
for modeling transmission of vibrational energy. 
Specifically, a model in acoustics that is close to those in electromagnetics can be found in the so-called statistical energy analysis (SEA) \cite{lyon2003}. 
This method is currently used as a key strategy in developing numerical tools for analysis of vibrational energy in complex mechanical systems \cite{2011TannerBook}.

Applications of RMT can be also found in the field of optics and photonics. As an example,  
chaotic cavities are used to create several irregular modes that eventually lock within a narrow localization bandwidth thus creating \emph{lasing}, whence the name 
``random laser'' \cite{Cao_1999,2006Stone}.
\section{Theory}
\label{sec:theory}
% Define impedance matrix in general
% Cite appendix here
% Statement of the RCM
The RCM is conveniently formulated in terms of an impedance or admittance matrix. The use of network theory for solving  
electromagnetic problems is widely accepted in engineering and physics \cite{Felsen_2007}. 
This often leads to theoretical formulae that can be efficiently computed and compared with actual 
measurements. 
In particular, the use of \emph{open-circuit} parameters such as impedances and admittances does not require any specific knowledge  
of the sources and detectors connected to the terminals where voltages and currents are supposed to exist. 
Imagine injecting an excitation \emph{current} into a small antenna radiating inside an irregular cavity. 
The multiple reflections and scattering of rays off the metallic (lossy) walls establish a reaction to the current density flowing 
on the antenna surface. This creates an electromotive force that builds up the response \emph{voltage} at the 
terminals of the antenna. In the case of a single port, the ratio of the reaction voltage over the excitation current defines the terminal impedance. 
This parameter has affinities with the concept of \emph{acoustic impedance}, defined by the ratio of the \emph{acoustic pressure}, analogous to the 
\emph{reaction voltage} in electromagnetic systems, and the \emph{acoustic volume flow}, analogous to the electrical excitation \emph{current}.
The acoustic impedance at a particular frequency indicates how much sound pressure is generated by a given air vibration at that frequency. 

In the specific case where the antenna radiated inside a cavity, the impedance is called the \emph{cavity impedance} 
\cite{Zheng2006a}. In the presence of multiple ports, the response effect of the excitation current on each 
antenna terminal can be easily separated by defining a cavity impedance matrix \cite{Zheng2006a}. 
\begin{equation}\label{eqn:z_cav_matrix}
 V_p = \sum_{p^{'}} Z^{cav}_{p p^{'}} \left ( k_0 \right ) I_{p^{'}} \,\, ,
\end{equation}
To find $Z_{p p^{'}}^{cav}$ in 3D cavities, a solution of the vector wave equation satisfying cavity boundary conditions is constructed. 
This is done through the usual procedure of expanding the cavity fields in a basis of electromagnetic and electrostatic eigenmodes. 
At this point, the prescriptions of the RCM based on wave chaos and random matrix theory are used. 
Firstly, the exact cavity spectrum is replaced with a spectrum of eigenvalues generated by a large random 
matrix of the GOE. The symmetry properties of the adopted matrix reflect the physical properties 
of the actual system, whence other ensembles can be invoked for generating the random eigenvalues, e.g., the Gaussian 
Unitary Ensemble (GUE) in case the time-reversal invariance is broken by the presence of a magnetized ferrite. 
Secondly, the mode amplitude (which is unknown) is replaced with a superposition of random plane-waves. Those two ingredients define what 
we call a ``chaotic mode'' of the irregular cavity.

Starting from a deterministic mode expansion, the RCM prescriptions lead to a fluctuating impedance matrix having dimension 
$N_p \times N_p$, where $N_p$ is the number of ports connected to the cavity, viz., 
\begin{equation}\label{eqn:Zcav_3D}
 \ushortdw{Z}^{cav} = i \Im \left \{ \ushortdw{Z}^{rad} \right \} + 
    \left [ \Re \left \{ \ushortdw{Z}^{rad} \right \} \right ]^{1/2} \cdot \ushortdw{\xi} \cdot \left [ \Re \left \{ \ushortdw{Z}^{rad} \right \} \right ]^{1/2} \,\, ,
\end{equation}
where $\ushortdw{Z}^{rad} = \ushortdw{R}^{rad} + i \Im \left ( \ushortdw{Z}^{rad} \right )$ is in the simplest theory an 
$N_p \times N_p$ diagonal matrix whose elements are the complex radiation impedances of each port.
Here, the radiation impedance provides the linear relation between voltages and currents at a port in the case in which 
waves are allowed to enter the enclosure through the port but not return, as if they were absorbed in the enclosure. 
The matrix $\ushortdw{\xi}$ naturally incorporates cavity homogeneous losses, and it is defined as 
\begin{equation}\label{eqn:universal_fluct}
 \ushortdw{\xi} = - \frac{i}{\pi} \sum_n \frac{ \underline{\Phi}_n \underline{\Phi}^{T}_n}{\mathcal{K}^2_0 - \mathcal{K}^2_n + i \alpha} \,\, ,
\end{equation}
where $\mathcal{K}^2_{\left ( \cdot \right )} = k^2_{\left ( \cdot \right )} / \Delta k^2$, and $\alpha$ is the loss parameter defined below. 
In the lossless case, $\ushortdw{\xi}$ is found to be an element of the Lorentzian ensemble \cite{Brouwer1997}.
Here, $\underline{\Phi}_n$ is a vector of uncorrelated, zero mean, unit width Gaussian random variables, and $k^2_{n}$ are the 
eigenvalues of a matrix selected from the GOE \cite{Mehta1991B}, where the central eigenvalue is 
shifted to be close to $k^2_0 = \omega^2 / c^2$, and $\omega$ is the frequency of excitation. 
%The shift implies that $\ushortdw{\xi}$ has zero mean. 
In case of high number of overlapping modes, e.g., occurring in overmoded reverberation chambers, the eigenvalues have uniform distribution 
and are distributed as per Wigner's ``semicircle law'' (see \cite[Appendix]{Hemmady2012} and \cite{Mehta1991B} for more details).
Eigenvalues associated with participating modes are scaled so that the average spacing between eigenvalues near the central
one is $\Delta k^2$, which is selected to match the mean spacing of resonances of the enclosure in the frequency range of interest. 
The effect of wall losses, internal homogeneous losses and the cavity dimensions (volume), are embedded into a single dimensionless loss parameter 
\begin{equation}\label{eqn:alpha_expl}
 \alpha = \frac{k^2_0}{Q \Delta k^2} \,\, ,
\end{equation}
appearing in (\ref{eqn:universal_fluct}), which is formally identical to the \emph{inverse finesse parameter}, widely used in optical 
resonators to quantify the granularity of resonances over a frequency band of interest \cite{born1999principles}, and where 
the (average) cavity quality factor is defined as 
\begin{equation}
 Q = \omega \,\frac{E}{P_d} \,\, ,
\end{equation}
with $E$ the average energy stored inside, and $P_d$ total power dissipated throughout the cavity.
The greater $\alpha$ the higher the resonance overlapping over the considered bandwidth. 

It is through the matrix $\ushortdw{\xi}$ that the propagation of waves in the enclosure from one port to another and back is modeled.
Such a propagation is universally fluctuating with statistics regulated by a single parameter, $\alpha$.
The RCM treats ports by means of a slowly-varying-in-frequency radiation impedance matrix.
The enclosure itself is modeled by (\ref{eqn:universal_fluct}) that has the physical meaning of a normalized impedance, modeling the chaotic scattering taking place throughout the cavity. 
Rapid variations of impedance with frequency arise from this chaotic behavior.
Similarly, in acoustics and vibrational systems, the acoustic impedance usually varies strongly when the frequency is changed.
The form (\ref{eqn:Zcav_3D}) makes a perfect link with well-established deterministic theories.
The novelty of the RCM resides in the definition of a `random' impedance that is \emph{dressed with deterministic features.} 

%integrate
In practical mode-stirred chambers, it is expected that perfect mixing leads to a zero mean complex field. However, an offset in the scatter plot of the measured field is often experienced. 
By adopting the Poisson kernel theory, the RCM has been extended to account for short-range ray trajectories (short-orbits) between the ports. 
Those short orbits create system-specific properties resulting in nonuniversal frequency fluctuations, and can be acconted for with an extension of the RMT. 
%As detailed in \cite{Zheng2006a, Zheng2006b, Hemmady2006T} the RCM separates the universal and the nonuniversal parts in terms of
%impedance matrix $\ushortdw{Z}$, or by a bilinear transform, in terms of scattering matrix $\ushortdw{S}$.
%The concept is similar to the Poisson kernel theory, but the advantage of the RCM is that nonuniversal contributions manifest themselves in $\ushortdw{Z}$ as a simple
%correction ``dressing'' the normalized impedance that is defined exclusively by the RMT \cite{Zheng2006a, Zheng2006b, Hart2009}.

The philosophy of the Poisson kernel \cite{Mello1985,Kuhl2005} is to include those scattering matrix contributions arising from parts of the system that are \emph{non-random}. 
This same perspective has been applied to the impedance matrix $\ushortdw{Z}$ of a microwave enclosure \cite{Hart2009}. 
Experiments involving mode-stirred cavities and ray-chaotic cavities often rely on the variation of boundary configuration to create several statistically independent realizations of the system 
and to compile statistics \cite{Kostas1991,Schafer2005_2,Kober2011}. However, imperfections in the realization of perfectly diffused fields have been observed \cite{2009_Moglie} numerically and 
experimentally. This resulted in the beginning of a serious effort to model and understanding these ``unstirred'' occurrences \cite{2010_Mariani, 2011_Ladbury}.  
From a wave chaotic perspective, it is noted that the channel-channel interaction introduces deterministic field components that remain fixed 
throughout the ensemble (such as wall proximity on scattering objects inside the enclosure) despite the design of a very efficient stirring structure \cite{2012_Moglie}. 
This kind of ``imperfection'' has been widely investigated in quantum chaos \cite{Prange2005, Bulgakov2006}. 
Therefore, relevant ray trajectories, which remain unchanged in many or all realizations of the ensemble, may exist and represent another system-specific feature. Hart \textit{et al.}
\cite{Hart2009, Yeh2010a, Yeh2010b} extended the RCM to take account of such ray trajectories that leave a port and soon return to it, or another port, instead of ergodically sampling the system. 
These ray trajectories are also called ``short orbits'' or ``short trajectories''. It is imagined that each ray leaves from a port $p$ at $\textbf{r}_i$ with a certain $\textbf{k}_i$ vector and it arrives 
at a final position $\textbf{r}_f$ with a wave vector $\textbf{k}_f$. It is further assumed that the ports are separated from the wall much more than a wavelength. 

The impedance form of the RCM is slightly modified by the presence of short-range trajectories. In particular, it is found by a semiclassical 
theory of orbits \cite{Hart2009}
\begin{equation}
  \ushortdw{Z}^{cav} = i \Im \left \{ \ushortdw{Z}^{rad} \right \} + 
    \ushortdw{v}^{\dagger} \cdot \frac{\ushortdw{1} + \ushortdw{T}}{\ushortdw{1} - \ushortdw{T}} \cdot \ushortdw{v}  \,\, ,
\end{equation} 
where $\ushortdw{T}$ is the Bogolomny \emph{transfer operator} which in the electromagnetic case is given by \cite{Bogomolny1992}
\begin{equation}
 T \left ( q_i, q_f, \textbf{k} \right ) = \frac{-i}{4} \sqrt{D_{\textbf{r}_i, \textbf{r}_f}} \, e^{i S \left ( \textbf{r}_i, \textbf{r}_i, \textbf{k} \right ) - i \frac{\pi}{4}} \, \sqrt{\frac{\cos \theta_f}{\cos \theta_i}} \,\, ,
\end{equation} 
where $\theta_i \left ( \theta_f \right )$ is the angle between the initial (final) wave vector and the surface at the position it leaves (hits), $S \left ( \textbf{r}_i, \textbf{r}_f, \textbf{k} \right )$ is the 
classic action along the direct trajectory from $\textbf{r}_i$ to $\textbf{r}_f$, and $ \sqrt{D_{\textbf{r}_i, \textbf{r}_f}}$ is the stability of the orbit from $\textbf{r}_i$ to $\textbf{r}_f$ (monodromy) 
that physically represents a geometrical factor of the trajectory taking account of the spreading of the ray tube along its path. It is also found that  
\begin{equation}
 \ushortdw{v}^{\dagger} \cdot \ushortdw{v} = \Re \left \{ \ushortdw{Z}^{rad} \right \} \,\, .
\end{equation}
Ultimately, the correction introduced by the presence of short-orbits in the RCM average impedance  as
\begin{equation}
 \left < \ushortdw{Z}^{cav} \right > = i \, \ushortdw{X}^{rad} + \left [ \ushortdw{R}^{rad} \right ]^{1/2} \cdot \ushortdw{z} \cdot \left [ \ushortdw{R}^{rad} \right ]^{1/2} \,\, ,
\end{equation}
where the elements of $\ushortdw{z}$ are expressed as 
\begin{equation}
 z_{n,m} = \sum_{b \left ( n, m \right )} \, - p_b \sqrt{D_b} \exp \left [ - \left ( i \left | \textbf{k} \right | + k^{'} \right ) L_b - i \left | \textbf{k} \right | L_{port} - i \beta_b \pi\right ] \,\, ,
\end{equation}
with $b \left ( n, m \right )$ an index over all classical trajectories which leave the $n^{th}$ port, bounce $\beta_b$ times, and return to the $m^{th}$ port, $L_b$ is the length of the trajectory $b$, 
$k^{'}= k / \left ( 2 Q \right )$ is the effective attenuation parameter taking account of loss through the average quality factor $Q$, and $L_{port}$ is the port-dependent constant length between the 
$n^{th}$ and $m^{th}$ port. 
The geometrical factor $D_b$ is now a function of the length of each segment of the trajectory, the angle of incidence of each bounce, and the radius of curvature of each wall encountered 
in that trajectory. 
The application of this theory to reverberation chambers could pave the way to understand and use the \emph{single-frequency} regime, where comparison of experimental and 
theoretical statistics often exhibits strong deviation from asymptotic distributions \cite{2009_Moglie}. 

%Previously, considerations related to the short-range trajectories arose in the field of quantum scattering theory tackled by a semiclassical approach \cite{Stockmann1999B,
%Gutzwiller1990B, Richter2002, Ishio1995, Prange2005, Stampfer2005}.
%The effect of short-orbits has been observed in a microwave billiard \cite{Zheng2006a, Zheng2006b, Zheng2006c} and for quantum transport in chaotic cavities \cite{Richter2002, Muller2007} and
%have either constrained or frustrated previous tests of RMT predictions. 
Researchers have examined short orbits in cases where the system and the ports can be treated in the semiclassical
approximation \cite{Prange2005} or considered the effect on eigenfunction correlations due to short orbits associated with
nearby walls \cite{Urbina2006}. 
Theoretical results have been previously obtained that acknowledge the effects of short-orbits in the context of quantum graphs \cite{Kottos2003}. 
The RCM for microwave cavities has been confirmed by experiments \cite{Hart2009, Yeh2010b}. Moreover, experiments in a
2D microwave cavity have extracted a measure of the microwave power that is emitted at a certain point in the cavity and returns to
the same point after following all possible classical trajectories of a given length \cite{Stein1992}. 
%Other experiments have been performed in the past. The Poisson Kernel approach can
%also be generalized to include short orbits \cite{Bulgakov2006} through measurement of a statistical optical $S$ matrix. 
None of this prior work developed a general first-principles deterministic
approach to experimentally analyze the effect of short ray trajectories.  
The effect of short trajectories in two-port wave-chaotic cavities has been demonstrated \cite{Yeh2010a, Yeh2010b} . 
The results of the short-orbit correction can be generalized to distributed ports and apertures by similar procedures.
For apertures, it is interesting to think about the proximity of a dielectric/metallic object that  is immersed in the wave-chaotic field. This physical situation has been investigated by 
Harrington in the case of regular cavities, where the aperture-object resonance was predicted by method of moments \cite{Harrington_1982}. 

\subsection{Aperture radiation of cavities}
\begin{figure}
   \centering
    \includegraphics[width=0.5 \textwidth]{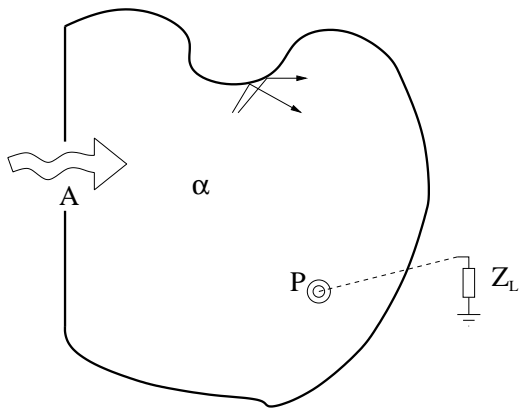}
    \caption{\label{fig:single_cav} Wave chaotic enclosure with both one aperture and one port.}
 \end{figure}

Besides localized ports, the RCM can also include a description of apertures, 
a situation depicted in Fig. \ref{fig:single_cav}.
% INSERT FIGURE OF CAVITY WITH APERTURE AND PORT
%The resulting model is believed to catalyze its use and validation through electromagnetic experiments. 
%Specifically, the main idea of including aperture radiation inside wave chaotic cavities has been followed and developed. 
Ongoing research is focusing on the extension to very large and irregular apertures, as well as to very small apertures, 
where singular fields strongly affect their (system specific) radiation \cite{Bethe_1944}. 
The aperture excitation of an irregular enclosure is indeed a rather new theoretical problem to investigate. 
Interestingly, such a theory develops well on the existing literature on \emph{cavity backed apertures}, and it links deterministic theories with statistical theories on reverberation chambers. 
The work of Harrington on cavities showed that boundary-value problems involving apertures are conveniently described in terms of the admittance matrix \cite{1976_Harrington}. 
Typically, radiation admittances are obtained by numerical computation of cavity eigenmodes. 
On the other hand, it is known that when the cavity is strongly perturbed by irregular objects and/or boundaries, modes change their topology and spectrum completely, 
e.g., mode stirred enclosures/chambers \cite{Bunting2004}, 
and the fields become amenable to statistical description \cite{hill2009}. 
Also in the presence of apertures, thanks to the RCM perspective of separating system-specific from universal characteristics, 
those two apparently distinct approaches can be unified through wave chaos theory. 
Similar to a port, the aperture geometry is fully included in the model through the radiation admittance matrix. 
The procedure to be followed is practically the same that has been used to arrive at Eq. (\ref{eqn:Zcav_3D}). A detailed mathematical derivation is described in \cite{Gradoni_EMCRoma_2012}, 
and is briefly recapitulated below.

The aperture has been treated as an opening in a zero thickness metallic plane, and the analysis started by expanding the aperture (tangential) field distribution in a basis of modes
\begin{equation}\label{eqn:Et_rad_aperture}
 \textbf{E}_t = \sum_s V_s \textbf{e}_s \left ( \textbf{x}_{\perp} \right ) \,\, ,
\end{equation} 
where $V_s$ are aperture voltages representing the electric mode amplitudes, $\textbf{x}_{\perp}$ is the spatial coordinate within the aperture area, $\textbf{e}_s$ are the aperture modes and have 
only transverse fields, normalized such that 
$\int_{aperture} d x^2_{\perp} \, \left | \textbf{e}_s \right |^2 = 1$. 
Solving Maxwell's equations with Dirichlet boundary conditions $\textbf{E}_t  =  0$, 
yields an expression for transverse magnetic fields in terms of the same basis of modes
\begin{equation}\label{eqn:Ht_rad_aperture}
 \textbf{H}_t = \sum_s I_s \hat{n} \times \textbf{e}_s \left ( \textbf{x}_{\perp} \right ) \,\, ,
\end{equation}
where $\hat{n}$ is the outward normal to the cavity, which has been taken to be in the $z$-direction. 
The linear relation between magnetic field mode amplitude $I_s$, and electric field mode amplitude $V_s$, is expressed in terms of an admittance
\begin{equation}\label{eqn:Y_aperture}
 I_s = \sum_{s^{'}} Y^{\left ( \cdot \right )}_{s s^{'}} \left ( k_0 \right ) V_{s^{'}} \,\, .
\end{equation}
Here $Y^{\left ( \cdot \right )}_{s s^{'}} = Y^{\left ( rad \right )}_{s s^{'}}$  or $Y^{\left ( \cdot \right )}_{s s^{'}} = Y^{\left ( cav \right )}_{s s^{'}}$ 
depending on whether the aperture radiates into free space or into a cavity. 
In particular, in presence of the cavity, the boundary-value problem is solved by expanding the cavity fields in modes, and then by applying the RCM prescriptions for the properties of these modes. 
In this case, the RCM formulation leads to a specific relation between free-space admittance $Y^{\left ( rad \right )}_{s s^{'}}$ and cavity admittance $Y^{\left ( cav \right )}_{s s^{'}}$, viz., 
\begin{equation}\label{eqn:Ycav_fluct}
\ushortdw{Y}^{cav} = i \Im \left \{ \ushortdw{Y}^{rad} \right \} + \left [ \Re \left \{ \ushortdw{Y}^{rad} \right \} \right ]^{1/2} \cdot 
 \ushortdw{\xi} \cdot \left [ \Re \left \{ \ushortdw{Y}^{rad} \right \} \right ]^{1/2} \,\, ,
\end{equation}
where the matrix $\ushortdw{\xi}$ is the same as defined in (\ref{eqn:universal_fluct}), and the elements of $\ushortdw{Y}^{rad}$ are defined by \cite{Gradoni_EMCRoma_2012}.
%For the off-diagonal elements of the radiation admittance, the following direct calculation can be performed 
%\begin{equation}\label{eqn:Y_Hilbert_slit}
% Y^{rad}_{n m} \left ( k_0 \right ) = G^{rad}_{n m} \left ( k_0 \right ) + \mathcal{H} \left [ G^{rad}_{n m} \left ( k \right ) \right ] \left ( k_0 \right ) + i \, B^{ms}_{n m} \left ( k_0 \right ) \,\, ,
%\end{equation}
%where $\mathcal{H} \left [ \cdot \right ]$ indicates the Hilbert transform.
%For diagonal elements with $n=m$, (\ref{eqn:Y_Hilbert_slit}) can be still used upon a proper correction that ensures the convergence of the Hilbert transform \cite{Gradoni_EMCRoma_2012}.
%It is to be noticed that the deterministic behavior of the aperture is retrieved on average, as $\left < \ushortdw{Y}^{cav} \right > = \ushortdw{Y}^{rad}$,
\subsection{Coupling with external radiation}
\begin{figure}[t]
\psfrag{t}{$\theta$}
\psfrag{x}{$\textbf{x}_{\perp}$}
\psfrag{k}{$\textbf{k}^{inc}$}
\psfrag{h}{$\textbf{h}^{inc}$}
\psfrag{n}{$\hat{z}$}
\psfrag{A}{$\mathcal{A}$}
\centering
\includegraphics[width=0.5 \textwidth]{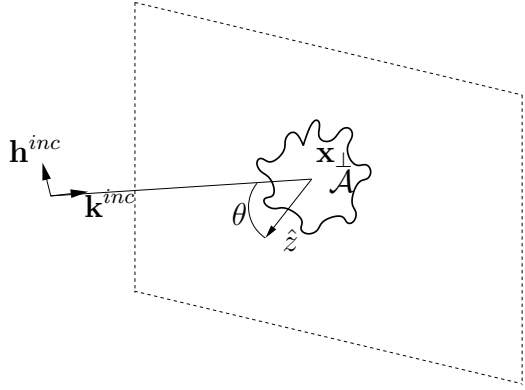}
\caption{\label{fig:ap_geo} Geometry of an aperture of arbitrary shaper illuminated by an external plane-wave.}
\end{figure}

In practical electromagnetic systems, an aperture can be fed by a waveguide with the same transverse sectional shape, or it can be exposed to external radiation. 
In both cases the tangential field expansions (\ref{eqn:Et_rad_aperture}) and (\ref{eqn:Ht_rad_aperture}) can be preserved. 
It is thus instructive to analyze the typical scenario depicted in Fig. \ref{fig:ap_geo} of an oblique plane-wave with wave vector $\textbf{k}^{inc}$ and polarization of 
magnetic field $\textbf{h}^{inc}$ (perpendicular to $\textbf{k}^{inc}$) exciting the aperture. In \cite{1976_Harrington}, this analysis has been performed by continuity of tangential 
magnetic field at the aperture plane, that leads to the equivalent network equation
% PUT FIGURE EXTERNAL RADIATION EXCITING APERTURE
by projecting the two magnetic field expressions on the aperture basis, and equating the amplitudes \cite{2012_GradoniA, 2012_GradoniB, Gradoni_EMCRoma_2012}
\begin{equation}\label{eqn:radiation_aperture}
 \left ( \ushortdw{Y}^{cav} + \ushortdw{Y}^{rad} \right ) \cdot \ushortw{V} = 2 \ushortw{I}^{inc} \,\, ,
\end{equation}
where $\ushortw{V}$ is a vector containing the voltages $V_s$, $I_s^{inc} = - \hat{n} \cdot \tilde{\textbf{e}}_s \left ( - \textbf{k}^{inc}_{\perp} \right ) \times \textbf{h}^{inc}$, and $\tilde{\textbf{e}}$ is the Fourier transform of the 
aperture basis mode for the electric field.
\subsection{Hybrid formulation: ports and apertures} 
The use and verification of the RCM for apertures through quadratic (power) measurements requires the presence of antennas/ports inside the cavity. 
Therefore, it becomes natural to formulate a hybrid RCM including both the aperture admittance matrix and the port impedance matrix.

The RCM has been extended to account for the joint presence of electrically wide openings of arbitrary shape and localized ports. 
A schematic framework of such a situation is reported in Fig. \ref{fig:single_cav}.
In that case, an input column vector $\ushortw{\phi}$ has been constructed that consists of the aperture voltages and terminal currents, and an output vector $\ushortw{\psi}$ 
consisting of the aperture currents and terminal voltages
 \begin{equation}
  \ushortw{\phi} = \left [ 
 \begin{array}{c}
    \ushortw{V}_{A}  \\
    \ushortw{I}_{P} 
 \end{array} \right ]
 \end{equation}
 and
 \begin{equation}
  \ushortw{\psi} = \left [ 
 \begin{array}{c}
    \ushortw{I}_{A}  \\
    \ushortw{V}_{P} 
 \end{array} \right ]
\end{equation} 
where $\ushortw{V}_{A,P}$ are the aperture and port voltages, and $\ushortw{I}_{A,P}$ are the aperture and port currents.
These are then related by a hybrid matrix $\ushortdw{H}$, $\ushortw{\psi} = \ushortdw{H} \cdot \ushortw{\phi}$, where
\begin{equation}\label{eqn:RCM_hybrid}
   \ushortdw{H} = i \, \Im \left ( \ushortdw{U} \right ) + \left [ \Re \left ( \ushortdw{U} \right ) \right ]^{1/2} \cdot 
   \ushortdw{\xi} \cdot \left [ \Re \left ( \ushortdw{U} \right ) \right ]^{1/2} \,\, .
\end{equation}
Here the matrices $\ushortdw{U}$ and $\ushortdw{V}$ are block diagonal, viz., 
\begin{equation}\label{eqn:matrix_U}
 \ushortdw{U} =  \left [ 
 \begin{array}{cc}
    \ushortdw{Y}^{rad} & 0  \\
    0 & \ushortdw{Z}^{rad} 
 \end{array} \right ] \,\, ,
\end{equation}
and
\begin{equation}
  \ushortdw{V} =  \Re \left ( \ushortdw{U} \right ) \,\, .
\end{equation}
The dimension of $\ushortdw{U}$ and $\ushortdw{V}$ is $\left ( N_s + N_p \right ) \times \left ( N_s + N_p \right )$, where $N_p$ is the number of port currents and 
$N_s$ is the number of aperture voltages. 
The linear problem at hand is thus formulated in terms of matrices with relatively small dimensions, 
typically given by the number of ports and aperture modes used to access the electromagnetic structure. 
Strictly speaking, the RCM formulation has facilitated a \emph{model order reduction} \cite{Felsen_2007} if compared to full-wave or other statistical models of cavity fields.
Usually the superposition of plane-waves or modes requires the computation of the field at each point of the structure, thus producing very large matrices. 
In the different perspective of looking at channels, the available number of degrees of freedom (typically constituted by the number of excited modes of the cavity) is reduced 
to the number of port profiles/aperture basis functions.
\section{Random coupling model in action: numerical simulations}
\label{sec:calculation}
\begin{figure}
\centering
\subfloat[]{\includegraphics[width=2.5in]{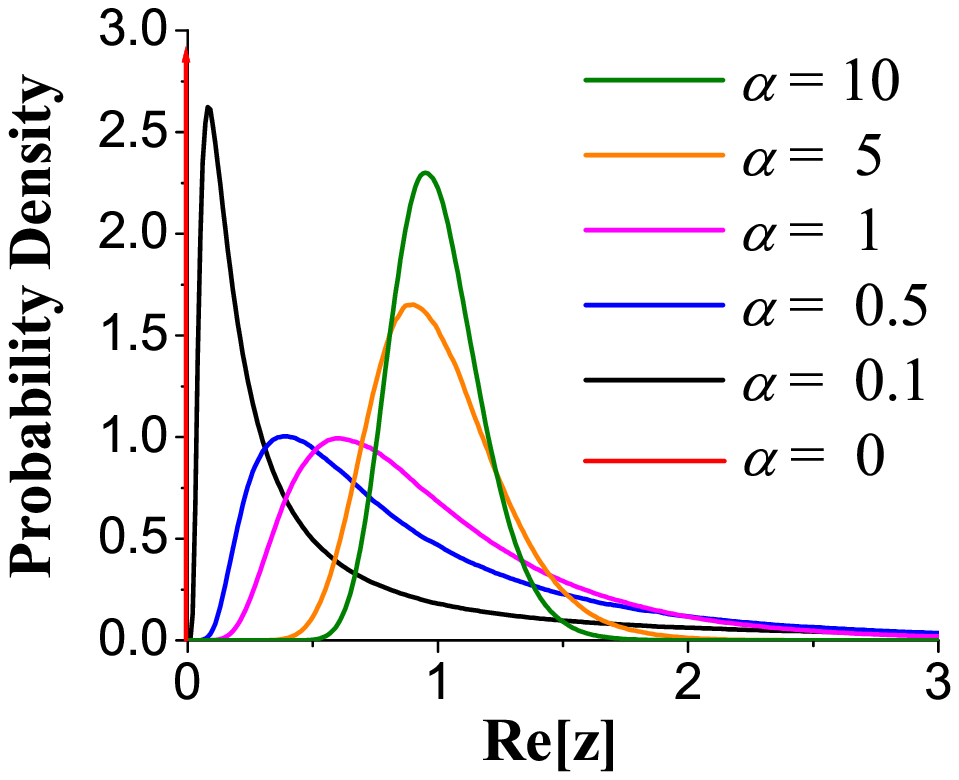}}
\subfloat[]{\includegraphics[width=2.5in]{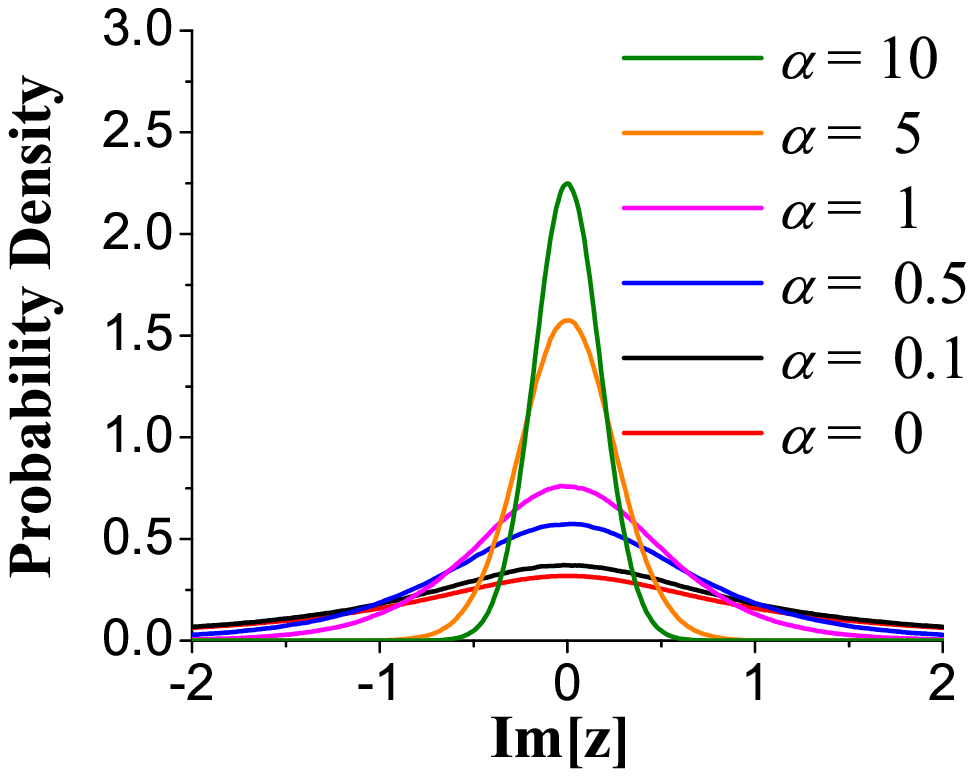}}
\caption{Random matrix theory prediction for the real (a) and the imaginary (b) part of the normalized impedance $z = \frac{Z^{cav} - i \Im \left \{ Z^{rad} \right \}}{\Re \left \{ Z^{rad} \right \}}$ 
 for various values of $\alpha$.}
\label{fig:norm_imp}
\end{figure}

Ensembles of cavities have been generated by the application of the Monte Carlo method described in \cite{Zheng2006a, Zheng2006b, Hemmady2012}. 
Once the system-specific properties of channels are predicted or measured separately under an $\alpha \rightarrow \infty$ (free-space) condition, Monte Carlo simulations of (\ref{eqn:universal_fluct}) 
allow for generating an ensemble of cavity impedances of the form (\ref{eqn:Zcav_3D}) or, identically, of cavity admittances of the form (\ref{eqn:Ycav_fluct}). 
A proper number of realizations must be chosen to create adequate statistics close to the asymptotic behavior, which is Lorentzian in the lossless regime ($\alpha = 0$), and Gaussian in the 
high-loss regime ($\alpha \gg 1$), see Fig. (\ref{fig:norm_imp}). First of all, the universally fluctuating impedances (\ref{eqn:universal_fluct}) is recast as
\begin{equation}\label{eqn:xi_MC}
  \ushortdw{\xi} = \frac{i}{\pi} \, \ushortdw{\Phi} \cdot \frac{\ushortdw{1}}{\ushortdw{\lambda} - i \alpha \ushortdw{1}} \cdot \ushortdw{\Phi}^{T} \,\, .
\end{equation}
The matrix $\ushortdw{\Phi}$ is a $M \times N$ coupling matrix with each element $\Phi_{i j}$ representing the coupling between the $i^{th}$ port profile/aperture mode ($1 \leq i \leq M$) and the $j^{th}$ 
chaotic mode of the cavity ($1 \leq j \leq N$). Each $\Phi_{i j}$ is an independent Gaussian-distributed random variable of zero mean and unit variance. The matrix $\ushortdw{\Phi}^{T}$ is the transpose 
of $\ushortdw{\Phi}$, and $\ushortdw{1}$ is a $N \times N$ identity matrix. The matrix $\ushortdw{\lambda}$ is an $N \times N$ diagonal matrix with a set of N random eigenvalues based on the 
GOE nearest-neighbor (Wigner) spacing distribution \cite{1967_Wigner}. 
By repeating this procedure many times in order to give a sufficiently large ensemble of $\ushortdw{Z}$, the statistical description of the cavity is thus generated.
Conversely, starting from cavity measurements of the impedance, the universal fluctuation of the single port impedance can be calculated by normalization of experimental results, to compare to RMT 
predictions of the universal behavior \cite{Hemmady2005a, Hemmady2005b}. 
Fig. \ref{fig:norm_imp} reports the distribution of the real (a) and the imaginary (b) part of the normalized impedance parametrized as a function of the loss factor $\alpha$. 
The loss parameter of the cavity at hand is predicted from the Weyl law on the basis of the volume of the cavity, its frequency of excitation ($\omega = k c$), 
and the electromagnetic losses. 
For a 3D enclosure it is found 
\begin{equation}\label{eqn:alpha3D}
 \alpha = \frac{k^3 V}{2 \pi^2 Q} \,\, ,
\end{equation}
where $V$ is the cavity volume, and $Q$ the average resonator quality factor.

The effect of increasing the loss parameter results in reducing the magnitude of fluctuation of impedance elements. 
It is to be noticed that for $\alpha = \infty$ we retrieve the case where ports/apertures radiate in free-space.
In the limit of infinitesimal losses the impedance fluctuation become very strong: this is a situation that can occur 
in superconducting cavities \cite{supercond_1998}, in low-loss optical systems, e.g., lasers \cite{2006Stone}, in microwave cavities operated in the \emph{non-Ericsson} (weak overlapping) regime 
\cite{Dietz2008, Dietz2009}, and in reverberation chambers/enclosures operated in the \emph{undermoded} regime \cite{2001_Arnaut,2003_Warne,2006_Picon,2011_Arnaut}.

In the limit of high losses, the diagonal (off-diagonal) elements of $\ushortdw{\xi}$ become unit (zero) mean Gaussian random variables.

\begin{figure}
\centering
\subfloat[]{\includegraphics[width=2.5in]{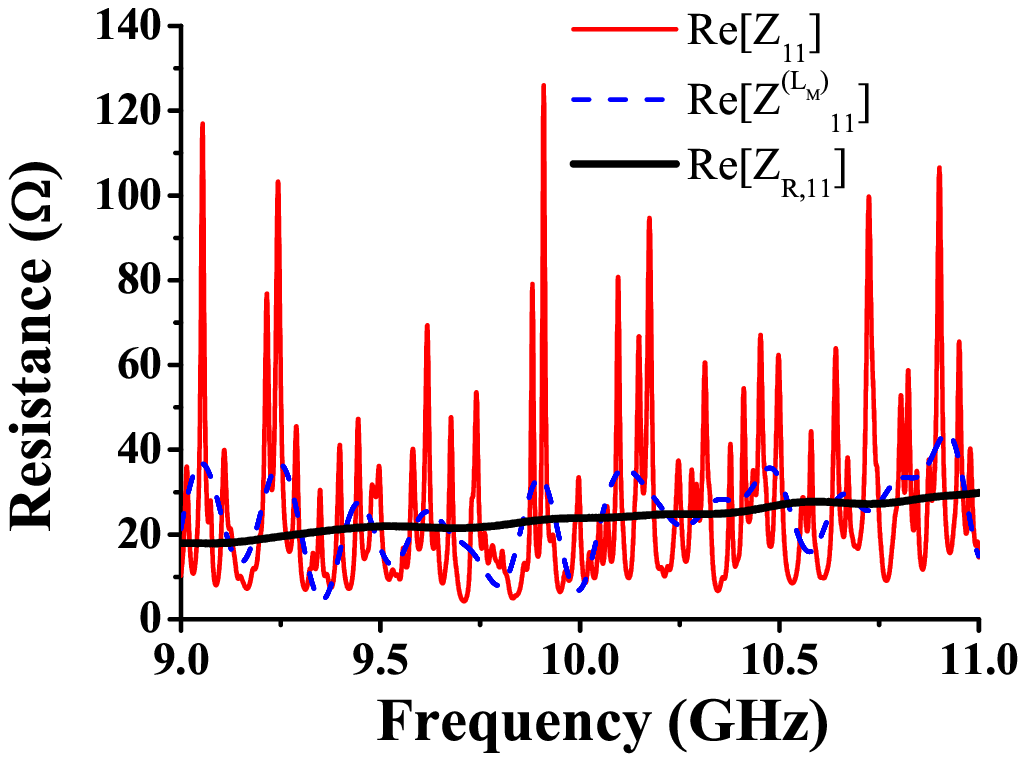}}
\subfloat[]{\includegraphics[width=2.5in]{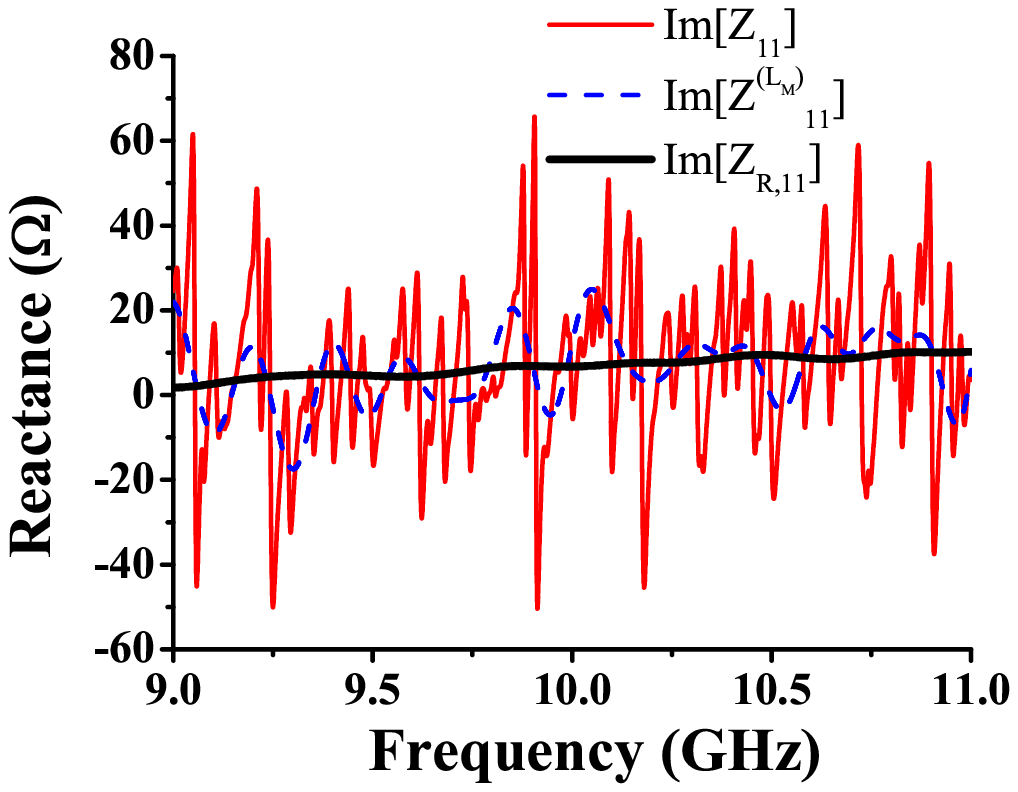}}
\caption{Resistance (a) and reactance (b) of the measured port (input) impedance in a ``bowtie'' cavity. The solid red curves are measured data. 
The dashed blue curves are created by combining the short orbit effect and the radiation impedance (thick black curves).}
\label{fig:short_orb_imp}
\end{figure}

Figs. \ref{fig:short_orb_imp} (red curves) reports a few experimental results for the frequency dependence 
of the real (a) and imaginary (b) single-port impedance measured at the coaxial port feeding a ``bowtie'' ray-chaotic cavity. 
A high frequency variability due to fluctuations that ride on top of several systematic trends can be noticed. 
Rapid fluctuations come from longer orbits.
The first systematic trend is shown as a black solid line and represents the measured radiation impedance of the port. 
The second systematic trend is shown as 
the thin dashed curves which are semi-numerical results that contain the radiation impedance (measured) and the effect of short (less than 200 cm) orbits.
This should be compared to the characteristic dimension of this billiard, which is $\mathcal{L}=\sqrt{A} \approx 34$ cm. Note that the short-orbit curve follows the major trend in the data, 
but does not describe the rapidly varying part of the impedance.
By removing all systematic effects up to an equivalent short orbit length of $\mathcal{L}_M = 200$ cm one can obtain a good correspondence between the theoretical and experimental 
normalized impedance \cite{Yeh2010b}.

The question now arises: how far should the short-orbit calculation be carried? 
At short times only a few of the shortest orbits will contribute, whereas the number of orbits contributing to the impedance will grow exponentially as the time increases. 
The relevant time scale for terminating the short-orbit calculation is the Ehrenfest time. 
The Ehrenfest time is a time scale associated with the classical-to-quantum crossover \cite{Aleiner1996}, and it is the time when wave packets start to deviate from the motion 
of deterministic classical ray trajectories to fully-developed wave chaos \cite{Jacquod2005}. 
Therefore, it is meaningless to compute the orbits whose lengths correspond to time scales much longer than the Ehrenfest time. 
For a microwave cavity, the Ehrenfest time can be calculated as 
\begin{equation}
 t_{E} = \frac{1}{h} \, \ln \left ( \frac{\mathcal{L}}{\lambda} \right ) \,\, , 
\end{equation}
where $h$ is the largest Lyapunov exponent of the classical ray dynamics in the cavity, 
$\mathcal{L}$ is the characteristic dimension of the cavity, and $\lambda$ is the wavelength of the probing waves. 
The Ehrenfest times of the bowtie cavity and the cut-circle cavity have been estimated. 
The Lyapunov exponent for the bowtie billiard is $h=0.7737$ $\textrm{ns}^{-1}$, so the Ehrenfest time at $f=10$ GHz is $3.2$ ns which corresponds 
to a length of $94$ cm in free space. For the cut-circle billiard we found $h=0.6411$ $\textrm{ns}^{-1}$, so the Ehrenfest time at $f=10$ GHz is $3$ ns which corresponds 
to a length of $92$ cm in free space.
\section{Interconnection of cavities} 
The RCM has been pushed further ahead to address the interconnection of complicated systems. 
Hence, the scenario of a linear chain of cavities has been investigated \cite{gradoni2012PRE}. 

\begin{figure}[t]
\centering
\includegraphics[width=0.8 \textwidth]{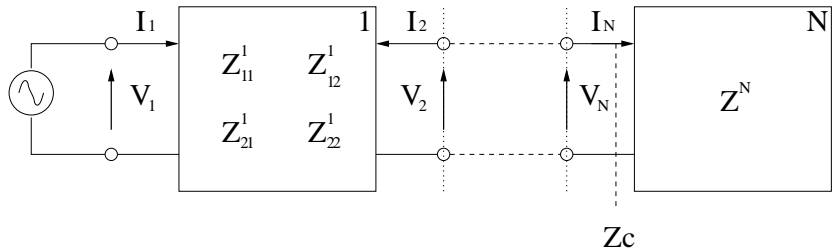}
\caption{\label{fig:cavity_chain}Network model of a chain of coupled cavities: the first cavity is excited by a localized source (port $1$), and the coupled field at the last cavity 
is picked up by a localized detector (port $N$). The cavity-to-cavity coupling can take place through either single-mode channels (transmission lines) or multimode channels (distributed ports or apertures).}
\end{figure}

The analyzed physical framework consists of $N$ interconnected two-port cavities, as shown in Fig. \ref{fig:cavity_chain}. 
% Insert figure of the linear chain of (Sinai) billiards
A fundamental quantity of interest for this analysis is the input impedance of the chain. 
This can be found in an iterated manner starting at the last cavity, where $Z^{(N)}_{in}=Z^{(N)}_{11}$, to the first cavity, to find $Z^{(1)}_{in}$
\begin{equation}
  Z^{(n)}_{in} = Z^{(n)}_{11} - \frac{\left ( Z^{(n)}_{12} \right )^2}{\left ( Z^{(n)}_{22} + Z^{(n+1)}_{in} \right )} \,\, .
\end{equation}
The concept of \emph{trans-impedance} has been adopted as a baseline to analyze and quantify the amount of signal coupled between 
two or more arbitrary elements (subsystems) of the chain. 
In particular, for the $n$-th cavity, the voltage at the first port of the $n+1$-st cavity $V^{(n+1)}_1$ is related to the current exciting the first port of 
the $n$-th cavity $I^{(n)}_1$ as
\begin{equation}\label{eqn:trans_imp}
 Z_T^{(n)} = \frac{V^{(n+1)}_1}{I^{(n)}_1} = \frac{Z^{(n+1)}_{in} Z^{(n)}_{21}}{Z^{(n)}_{22} + Z^{(n+1)}_{in}} \,\, .
\end{equation}
Consequently, the ratio of power entering the $N$-th cavity to that entering the first cavity can be expressed as a product, viz.,
\begin{equation}
 R_N \equiv \frac{P^{(N)}_{in}}{P^{(1)}_{in}} \,\, , 
\end{equation}
where each impedance involved in (\ref{eqn:trans_imp}) is expressed by the RCM.

\begin{figure}[t]
\centering
\includegraphics[width=0.65 \textwidth]{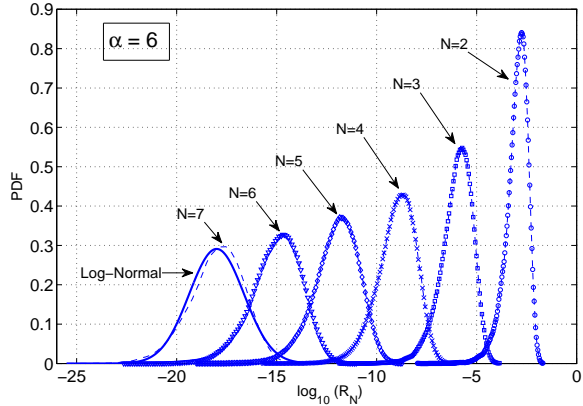}
\caption{\label{fig:logPratio_a6_7cav} Probability density function of power ratios $R_N \left ( \alpha=6 \right ) \equiv \frac{P^{(N)}_{in}}{P^{(1)}_{in}}$ for chains of up to seven cavities: 
the high-loss regime is assumed for all the (statistically identical) cavities in the chain.}
\end{figure}

The case of statistically identical cavities having moderate/high losses $\alpha_1 = \alpha_2 = \ldots = \alpha_N = \alpha > 1$, deserves special attention. 
Correspondingly, the weak fluctuation approximation can be employed leading to a simplified expression of (\ref{eqn:trans_imp}). 
More precisely, in the high-loss regime, $Z^{(n)}_{in} \approx Z^{(n)}_{11}$, and fluctuating impedances can be tackled with first-order 
perturbation theory. This procedure results in a simplification for the denominator of (\ref{eqn:trans_imp}), and
leads to an expression for the coupled power ratio $R_N$ that factorizes
\begin{equation}\label{eqn:RN_high_loss}
 R_N \left ( \alpha \right ) \approx \prod_{n=1}^{N-1} \, x^{(n)} \, \frac{t^{\left ( n, n+1 \right )}}{2 \pi \alpha} \,\, , 
\end{equation}
where $x^{(n)}$ are exponentially distributed random variabes
\begin{equation}\label{eqn:prod_chain_df}
 f_X \left ( x \right ) = \exp \left ( - x \right ) \,\, .
\end{equation}
The transmission factors in (\ref{eqn:RN_high_loss}) are given by
\begin{equation}\label{eqn:transm_chain}
 t^{\left ( n, n+1 \right )} = \frac{R^{(n)}_{22, rad} \, R^{(n+1)}_{11, rad}}{\left | Z^{(n)}_{22, rad} + Z^{(n+1)}_{11, rad} \right |^{2}} \,\, ,
\end{equation}
modeling the coupling between the $n$-th and $n+1$-st cavity, and the loss factor is given by (\ref{eqn:alpha3D}).
In \cite{gradoni2012PRE}, the Monte Carlo method described in section \ref{sec:calculation} has been used to generate ensembles of impedance matrices and compute the coupling in chains of cavities. 
Basically, the impedance matrix for each single cavity has been generated, and used in the theoretical expressions (\ref{eqn:Zcav_3D}) and (\ref{eqn:xi_MC}). 
Then, the power ratio has been evaluated for sequences of cavity chains of different length. For these studies, the cavities were assumed to be statistically identical. 
The PDFÕs of the logarithm of the power ratio for the case $\alpha=6$ is shown in Fig. \ref{fig:logPratio_a6_7cav}. 
%INSERT FIGURE POWER CHAIN
The high loss case (Fig. \ref{fig:logPratio_a6_7cav}) is well approximated by the analytic formulas (\ref{eqn:RN_high_loss}), (\ref{eqn:prod_chain_df}) and (\ref{eqn:transm_chain}). 
The distribution approaches log-normal as the number of cavities becomes large.  This is evident from the comparison of the $7$-cavity chain with a best-fit log-normal distribution. 
This behavior is expected since according to (\ref{eqn:RN_high_loss}) and (\ref{eqn:prod_chain_df}) the power ratio becomes a product of $N$ independent and identically distributed random variables, 
and thus, as $N$ becomes large the logarithm of the power ratio will be normally distributed. 
Anderson localization in the lossless case has been observed with respect to the number of chain elements by adopting a cascaded RCM in a chaotic dynamics perspective \cite{gradoni2012PRE}. 
\subsection{Connection with acoustics and vibrations}
\begin{figure}[t]
\psfrag{a}{$cavity$}
\psfrag{e1}{$E_i$}
\psfrag{e2}{$E_j$}
\psfrag{d1}{$\overline{d}_i$}
\psfrag{d2}{$\overline{d}_j$}
\psfrag{p1}{$P_i$}
\psfrag{p2}{$P_j$}
\psfrag{t}{$T^{\left ( i j \right )} $}
\centering
\includegraphics[width=0.65 \textwidth]{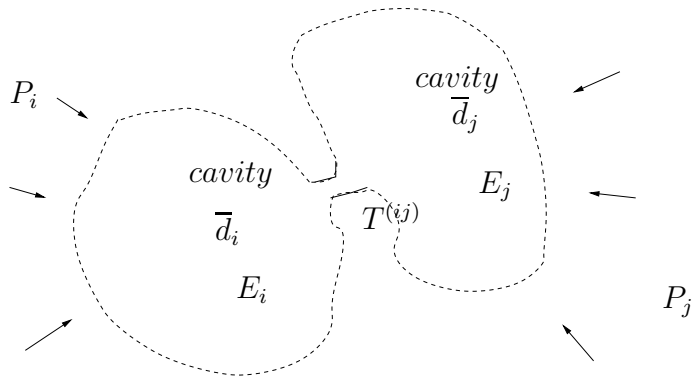}
\caption{\label{fig:chain_section} Arbitrary section of a chain: two cavities are excited by different input powers and they are weakly coupled. 
This simplified framework has been used to calculate the net power exchanged by the two cavities. A similar attack is adopted by the statistical energy analysis (SEA) approach in acoustics.}
\end{figure}

It turns out that the flow of electromagnetic energy can be expressed in a similar way to that used to model the flow of vibrational wave energy. 
The global system is divided into complex subsystems interconnected with each other through simple deterministic structures.
In particular, \cite{lyon2003} applied a ``thermodynamical'' approach to arrive at a statistical model of energy exchange between two parts (subsystems) 
of the system, namely statistical energy analysis (SEA). 
Specifically, by using energy \emph{equipartition} arguments, the average power passing from subsystem $i$ to subsystem $j$ ($P_{ij}$) can be expressed as \cite[Eq. (92)]{2007_Tanner}  
\begin{equation}\label{eqn:SEA_vibr}
 P_{ij} = \omega \overline{d}_i \eta_{ij} \left ( \frac{E_i}{\overline{d}_i} - \frac{E_j}{\overline{d}_j} \right ) \,\, ,
\end{equation}
where $\omega$ is the mean frequency of the source, $\eta_{ij}$ is the coupling loss factor, $\overline{d}_{(\cdot)}$ is the mean density of eigenfrequencies (modes) of the uncoupled 
subsystems, and $E_{( \cdot )}$ is the energy stored in each subsystem. 
In a similar way, by recalling the average power ratio (\ref{eqn:RN_high_loss}), it is worth noticing that the difference between the power entering two arbitrary elements of the chain can be written 
in the very general form 
\begin{equation}\label{eqn:RCM_SEM_affinity}
 P_{ij} = \frac{P_i \, T^{\left ( i, j \right )}}{2 \pi \alpha^{(i)}} - \frac{P_j \, T^{\left ( i, j \right )}}{2 \pi \alpha^{(j)}} \,\, ,
\end{equation}
valid for an arbitrary chain section of two elements, 
In (\ref{eqn:RCM_SEM_affinity}), $P_i$ is the input power entering cavity $i$, and $P_j$ is the input power entering cavity $j$. 
In equilibrium, the net power exchange between the to subsystems is established by the power dissipated in each cavity, and by the fraction of power 
leaking through the small (deterministic) channel.
The scheme of the physical framework analyzed to derive (\ref{eqn:RCM_SEM_affinity}) is reported in Fig. \ref{fig:chain_section}.
% PUT scheme of two coupled cavities for affinity with SEA
By using (\ref{eqn:alpha_expl}) the loss factor $\alpha^{(\cdot)}$, and by assuming  
\begin{equation}\label{eqn:dissipation_factor}
 \frac{P_i}{P^{(i)}_d} \approx \frac{P_j}{P^{(j)}_d} = L\,\, ,
\end{equation}
where $L$ is a dissipation factor of the subsystem since $P^{(\cdot)}_d$ is the dissipated power within each cavity, it is found that 
\begin{equation}\label{eqn:SEA_em}
 P_{ij} \approx \omega \frac{L}{2 \pi} T^{\left ( i j \right )} \left ( \frac{E_i}{\overline{d}_i} - \frac{E_j}{\overline{d}_j} \right ) \,\, ,
\end{equation}
where a coupling loss factor $\eta_{i j} = L T^{\left ( i j \right )} / 2 \pi$ is recognized, and 
\begin{equation}
 \overline{d}_{\left ( \cdot \right )} = \frac{k^2}{\Delta k^2_{\left ( \cdot \right )}} \,\, ,
\end{equation} 
is the mean (mode) density of each subsystem.
The assumption in (\ref{eqn:dissipation_factor}) is related to the equipartition of energy. This results in (\ref{eqn:SEA_em}) being formally similar 
to (\ref{eqn:SEA_vibr}) even in the case of autonomous (sub)systems coupled by an electrically short transmission line. 
Interestingly, in deriving the terms of (\ref{eqn:SEA_em}), a weak coupling argument has been used \cite{gradoni2012PRE} that is identical to  
the one typically used in SEA \cite{1990_Langley}. 
Methods based on SEA have been recently integrated with the finite element method (FEM) to tackle very complex vibro-acoustic scenarios \cite{2010_Tanner}.
\section{Conclusion and future perspective}
The general ``dressed'' impedance obtained in formulating the RCM for ports and terminals radiating inside complex electromagnetic cavities has been described. 
This statistical model accounts for short orbit deviations from the full chaotic mixing assumption. 
Numerical and experimental results validate the ``dressing'' of a universal fluctuating impedance matrix (dependent on a single loss parameter) with 
radiation impedance/admittance matrices (modeling radiation of ports/apertures). A generalization of the RCM to the scenario of interconnected cavities 
for weak coupling conditions is presented, and the distribution of the power flowing through such a chain discussed for an increasing number of cavities.
The analogy with Statistical Energy Analysis used in acoustics is pointed out by formulating a new model for the power exchanged by two wave chaotic cavities 
excited by different input powers.
\section*{Acknowledgement}
Work supported by the Air Force Office of Scientific Research grant FA95501010106 and the Office of Naval Research grant N000140911190.
\bibliographystyle{elsarticle-harv}
\bibliography{reference}

\begin{thebibliography}{87}
\expandafter\ifx\csname natexlab\endcsname\relax\def\natexlab#1{#1}\fi
\expandafter\ifx\csname url\endcsname\relax
  \def\url#1{\texttt{#1}}\fi
\expandafter\ifx\csname urlprefix\endcsname\relax\def\urlprefix{URL }\fi

\bibitem[{Aleiner and Larkin(1996)}]{Aleiner1996}
Aleiner, I.~L., Larkin, A.~I., Nov 1996. Divergence of classical trajectories
  and weak localization. Phys. Rev. B 54, 14423--14444.

\bibitem[{Alhassid(2000)}]{2000_Alhassid}
Alhassid, Y., Oct 2000. The statistical theory of quantum dots. Rev. Mod. Phys.
  72, 895--968.
\newline\urlprefix\url{http://link.aps.org/doi/10.1103/RevModPhys.72.895}

\bibitem[{Alt et~al.(1998)Alt, B\"acker, Dembowski, Gr\"af, Hofferbert,
  Rehfeld, and Richter}]{supercond_1998}
Alt, H., B\"acker, A., Dembowski, C., Gr\"af, H.-D., Hofferbert, R., Rehfeld,
  H., Richter, A., Aug 1998. Mode fluctuation distribution for spectra of
  superconducting microwave billiards. Phys. Rev. E 58, 1737--1742.
\newline\urlprefix\url{http://link.aps.org/doi/10.1103/PhysRevE.58.1737}

\bibitem[{Antonsen et~al.(2011)Antonsen, Gradoni, Anlage, and
  Ott}]{Antonsen_2011}
Antonsen, T.~M., Gradoni, G., Anlage, S., Ott, E., August 2011. Statistical
  characterization of complex enclosures with distributed ports. In:
  Proceedings of the IEEE International Symposium on EMC. Long Beach, CA (USA).

\bibitem[{Arnaut(2001)}]{2001_Arnaut}
Arnaut, L., nov 2001. Operation of electromagnetic reverberation chambers with
  wave diffractors at relatively low frequencies. Electromagnetic
  Compatibility, IEEE Transactions on 43~(4), 637 --653.

\bibitem[{Arnaut(2003)}]{2003_Arnaut}
Arnaut, L., feb 2003. Statistics of the quality factor of a rectangular
  reverberation chamber. Electromagnetic Compatibility, IEEE Transactions on
  45~(1), 61 -- 76.

\bibitem[{Arnaut and Gradoni(2011)}]{2011_Arnaut}
Arnaut, L., Gradoni, G., aug. 2011. On distributions of fields and power in
  undermoded mode-stirred reverberation chambers. In: General Assembly and
  Scientific Symposium, 2011 XXXth URSI. pp. 1 --4.

\bibitem[{Baranger and Mello(1996)}]{Baranger1996}
Baranger, H.~U., Mello, P.~A., 1996. Short paths and information theory in
  quantum chaotic scattering: transport through quantum dots. Europhys. Lett.
  33, 465.

\bibitem[{Beenakker(1997)}]{Beenakker1997}
Beenakker, C. W.~J., 1997. Random-matrix theory of quantum transport. Rev. Mod.
  Phys. 69, 731.

\bibitem[{Berry(1977)}]{Berry1977}
Berry, M.~V., 1977. Regular and irregular semiclassical wavefunctions. J. Phys.
  A 10, 2083.

\bibitem[{Bethe(1944)}]{Bethe_1944}
Bethe, H.~A., 1944. Theory of diffraction by small holes. Phys.\ Rev. 66~(7),
  163--182.

\bibitem[{Bogomolny(1992)}]{Bogomolny1992}
Bogomolny, E.~B., 1992. Semiclassical quantization of multidimensional systems.
  Nonlinearity 5, 805.

\bibitem[{Born and Wolf(1999)}]{born1999principles}
Born, M., Wolf, E., 1999. Principles of Optics: Electromagnetic Theory of
  Propagation, Interference and Diffraction of Light. Cambridge University
  Press.
\newline\urlprefix\url{http://books.google.com/books?id=oV80AAAAIAAJ}

\bibitem[{Brouwer(1995)}]{Brouwer1995}
Brouwer, P.~W., 1995. Generalized circular ensemble of scattering matrices for
  a chaotic cavity with nonideal leads. Phys. Rev. B 51, 16878.

\bibitem[{Brouwer and Beenakker(1997)}]{Brouwer1997}
Brouwer, P.~W., Beenakker, C. W.~J., 1997. Voltage-probe and
  imaginary-potential models for dephasing in a chaotic quantum dot. Phys. Rev.
  B 55, 4695.

\bibitem[{Bulgakov et~al.(2006)Bulgakov, Gopar, Mello, and
  Rotter}]{Bulgakov2006}
Bulgakov, E.~N., Gopar, V.~A., Mello, P.~A., Rotter, I., 2006. Statistical
  study of the conductance and shot noise in open quantum-chaotic cavities:
  Contribution from whispering gallery modes. Phys. Rev. B 73, 155302.

\bibitem[{Bunting and Yu(2004)}]{Bunting2004}
Bunting, C., Yu, S.-P., may 2004. Field penetration in a rectangular box using
  numerical techniques: an effort to obtain statistical shielding
  effectiveness. Electromagnetic Compatibility, IEEE Transactions on 46~(2),
  160 -- 168.

\bibitem[{Cao et~al.(1999)Cao, Zhao, Ho, Seelig, Wang, and Chang}]{Cao_1999}
Cao, H., Zhao, Y.~G., Ho, S.~T., Seelig, E.~W., Wang, Q.~H., Chang, R. P.~H.,
  Mar 1999. Random laser action in semiconductor powder. Phys. Rev. Lett. 82,
  2278--2281.

\bibitem[{Couchman et~al.(1992)Couchman, Ott, and Antonsen}]{Couchman_1992}
Couchman, L., Ott, E., Antonsen, T.~M., Nov 1992. Quantum chaos in systems with
  ray splitting. Phys. Rev. A 46, 6193--6210.

\bibitem[{Dietz et~al.(2009)Dietz, Friedrich, Harney, Miski-Oglu, Richter,
  Sch\"{a}fer, Verbaarschot, and Weidenm\"{u}ller}]{Dietz2009}
Dietz, B., Friedrich, T., Harney, H.~L., Miski-Oglu, M., Richter, A.,
  Sch\"{a}fer, F., Verbaarschot, J., Weidenm\"{u}ller, H.~A., 2009. Induced
  violation of time-reversal invariance in the regime of weakly overlapping
  resonances. Phys. Rev. Lett. 103, 064101.

\bibitem[{Dietz et~al.(2008)Dietz, Friedrich, Harney, Miski-Oglu, Richter,
  Sch\"{a}fer, and Weidenm\"{u}ller}]{Dietz2008}
Dietz, B., Friedrich, T., Harney, H.~L., Miski-Oglu, M., Richter, A.,
  Sch\"{a}fer, F., Weidenm\"{u}ller, H.~A., 2008. Chaotic scattering in the
  regime of weakly overlapping resonances. Phys. Rev. E 78, 055204.

\bibitem[{Doron et~al.(1990)Doron, Smilansky, and Frenkel}]{Doron1990}
Doron, E., Smilansky, U., Frenkel, A., 1990. Experimental demonstration of
  chaotic scattering of microwaves. Phys. Rev. Lett. 65, 3072.

\bibitem[{Felsen et~al.(2007)Felsen, Mongiardo, and Russer}]{Felsen_2007}
Felsen, L.~B., Mongiardo, M., Russer, P., 2007. Electromagnetic Field
  Computation by Network Methods. Springer.

\bibitem[{Fyodorov and Savin(2004)}]{Fyodorov2004}
Fyodorov, Y.~V., Savin, D.~V., 2004. Statistics of impedance, local density of
  states, and reflection in quantum chaotic systems with absorption. JETP Lett.
  80, 725.

\bibitem[{Fyodorov et~al.(2005)Fyodorov, Savin, and Sommers}]{Fyodorov2005}
Fyodorov, Y.~V., Savin, D.~V., Sommers, H.-J., 2005. Scattering, reflection and
  impedance of waves in chaotic and disordered systems with absorption. J.
  Phys. A: Math. Gen. 38, 10731.

\bibitem[{Gradoni et~al.(2012{\natexlab{a}})Gradoni, Antonsen, Anlage, and
  Ott}]{2012_GradoniB}
Gradoni, G., Antonsen, T., Anlage, S., Ott, E., sept. 2012{\natexlab{a}}.
  Coupling of external radiation to circuitry inside complex em environments.
  In: Electromagnetics in Advanced Applications (ICEAA), 2012 International
  Conference on. pp. 1233 --1234.

\bibitem[{Gradoni et~al.(2012{\natexlab{b}})Gradoni, Antonsen, Anlage, and
  Ott}]{2012_GradoniA}
Gradoni, G., Antonsen, T., Anlage, S., Ott, E., sept. 2012{\natexlab{b}}.
  External radiation of complex cavities described by the random coupling
  model. In: Electromagnetics in Advanced Applications (ICEAA), 2012
  International Conference on. pp. 357 --358.

\bibitem[{Gradoni et~al.(2012{\natexlab{c}})Gradoni, Antonsen, Anlage, and
  Ott}]{Gradoni_EMCRoma_2012}
Gradoni, G., Antonsen, T.~M., Anlage, S., Ott, E., September
  2012{\natexlab{c}}. Theoretical analysis of apertures radiating inside wave
  chaotic cavities. In: Proceedings of the EMC Europe. Rome (Italy).

\bibitem[{Gradoni et~al.(2012{\natexlab{d}})Gradoni, Antonsen, and
  Ott}]{gradoni2012PRE}
Gradoni, G., Antonsen, T.~M., Ott, E., Oct 2012{\natexlab{d}}. Impedance and
  power fluctuations in linear chains of coupled wave chaotic cavities. Phys.
  Rev. E 86, 046204.
\newline\urlprefix\url{http://link.aps.org/doi/10.1103/PhysRevE.86.046204}

\bibitem[{Gr\"af et~al.(1992)Gr\"af, Harney, Lengeler, Lewenkopf,
  Rangacharyulu, Richter, Schardt, and Weidenm\"uller}]{1992_graf}
Gr\"af, H.-D., Harney, H.~L., Lengeler, H., Lewenkopf, C.~H., Rangacharyulu,
  C., Richter, A., Schardt, P., Weidenm\"uller, H.~A., Aug 1992. Distribution
  of eigenmodes in a superconducting stadium billiard with chaotic dynamics.
  Phys. Rev. Lett. 69, 1296--1299.
\newline\urlprefix\url{http://link.aps.org/doi/10.1103/PhysRevLett.69.1296}

\bibitem[{Guhr et~al.(1998)Guhr, M\"{u}ller-Groeling, and
  Weidenm\"{u}ller}]{Guhr1998}
Guhr, T., M\"{u}ller-Groeling, A., Weidenm\"{u}ller, H.~A., 1998. Random-matrix
  theories in quantum physics: common concepts. Phys. Rep. 299, 189.

\bibitem[{Harrington(1982)}]{Harrington_1982}
Harrington, R., mar 1982. Resonant behavior of a small aperture backed by a
  conducting body. Antennas and Propagation, IEEE Transactions on 30~(2), 205
  -- 212.

\bibitem[{Harrington and Mautz(1976)}]{1976_Harrington}
Harrington, R.~F., Mautz, J.~R., 1976. A generalized network formulation for
  aperture problems. IEEE Transactions on Antennas and Propagation 24~(6), 870
  -- 873.

\bibitem[{Hart et~al.(2009)Hart, Antonsen, and Ott}]{Hart2009}
Hart, J.~A., Antonsen, T.~M., Ott, E., 2009. Effect of short ray trajectories
  on the scattering statistics of wave chaotic systems. Phys. Rev. E 80,
  041109.

\bibitem[{Hemmady et~al.(2012)Hemmady, Antonsen, Ott, and Anlage}]{Hemmady2012}
Hemmady, S., Antonsen, T.~M., Ott, E., Anlage, S.~M., 2012. Statistical
  prediction and measurement of induced voltages on components within
  complicated enclosures: A wave-chaotic approach. IEEE Trans. Electromag.
  Compat. 99, 1.

\bibitem[{Hemmady et~al.(2005{\natexlab{a}})Hemmady, Zheng, Antonsen, Ott, and
  Anlage}]{Hemmady2005b}
Hemmady, S., Zheng, X., Antonsen, T.~M., Ott, E., Anlage, S.~M.,
  2005{\natexlab{a}}. Universal statistics of the scattering coefficient of
  chaotic microwave cavities. Phys. Rev. E 71, 056215.

\bibitem[{Hemmady et~al.(2006{\natexlab{a}})Hemmady, Zheng, Antonsen, Ott, and
  Anlage}]{Hemmady2006b}
Hemmady, S., Zheng, X., Antonsen, T.~M., Ott, E., Anlage, S.~M.,
  2006{\natexlab{a}}. Aspects of the scattering and impedance properties of
  chaotic microwave cavities. Acta Physica Polonica A 109, 65.

\bibitem[{Hemmady et~al.(2006{\natexlab{b}})Hemmady, Zheng, Antonsen, Ott, and
  Anlage}]{Hemmady2006a}
Hemmady, S., Zheng, X., Antonsen, T.~M., Ott, E., Anlage, S.~M.,
  2006{\natexlab{b}}. Universal properties of two-port scattering, impedance,
  and admittance matrices of wave-chaotic systems. Phys. Rev. E 74, 036213.

\bibitem[{Hemmady et~al.(2005{\natexlab{b}})Hemmady, Zheng, Ott, Antonsen, and
  Anlage}]{Hemmady2005a}
Hemmady, S., Zheng, X., Ott, E., Antonsen, T.~M., Anlage, S.~M.,
  2005{\natexlab{b}}. Universal impedance fluctuations in wave chaotic systems.
  Phys. Rev. Lett. 94, 014102.

\bibitem[{Hill(2009)}]{hill2009}
Hill, D., 2009. Electromagnetic Fields in Cavities: Deterministic and
  Statistical Theories. IEEE Press Series on Electromagnetic Wave Theory.
  Wiley.
\newline\urlprefix\url{http://books.google.com/books?id=6BhWMXBf3JYC}

\bibitem[{Holland and John(1999)}]{Holland1999B}
Holland, R., John, R.~S., 1999. Statistical Electromagnetics. Taylor and
  Francis.

\bibitem[{Ishio and Burgd\"{o}rfer(1995)}]{Ishio1995}
Ishio, H., Burgd\"{o}rfer, J., 1995. Quantum conductance fluctuations and
  classical short-path dynamics. Phys. Rev. B 51, 2013.

\bibitem[{K\"ober et~al.(2011)K\"ober, Kuhl, St\"ockmann, Goussev, and
  Richter}]{Kober2011}
K\"ober, B., Kuhl, U., St\"ockmann, H.-J., Goussev, A., Richter, K., Jan 2011.
  Fidelity decay for local perturbations: Microwave evidence for oscillating
  decay exponents. Phys. Rev. E 83, 016214.

\bibitem[{Kostas and Boverie(1991)}]{Kostas1991}
Kostas, J.~G., Boverie, B., 1991. Statistical model for a mode-stirred chamber.
  IEEE Trans. EMC 33, 366.

\bibitem[{Kottos and Smilansky(2003)}]{Kottos2003}
Kottos, T., Smilansky, U., 2003. Quantum graphs: a simple model for chaotic
  scattering. J. Phys. A: Math. Gen. 36, 3501.

\bibitem[{Kuhl et~al.(2005)Kuhl, Mart\'{i}nez-Mares, M\'{e}ndez-S\'{a}nchez,
  and St\"{o}ckmann}]{Kuhl2005}
Kuhl, U., Mart\'{i}nez-Mares, M., M\'{e}ndez-S\'{a}nchez, R.~A., St\"{o}ckmann,
  H.~J., 2005. Direct processes in chaotic microwave cavities in the presence
  of absorption. Phys. Rev. Lett. 94, 144101.

\bibitem[{Langley(1990)}]{1990_Langley}
Langley, R., 1990. A derivation of the coupling loss factors used in
  statistical energy analysis. Journal of Sound and Vibration 141~(2), 207 --
  219.

\bibitem[{Lyon(2003)}]{lyon2003}
Lyon, R., 2003. Statistical Energy Analysis of Dynamical Systems: Theory and
  Applications. Mit Press.
\newline\urlprefix\url{http://books.google.com/books?id=hHQRPwAACAAJ}

\bibitem[{Maksimov and Tanner(2011)}]{2011TannerBook}
Maksimov, D., Tanner, G., 2011. A New Hybrid Method to Predict the Distribution
  of Vibro-Acoustic Energy in Complex Built-Up Structures. Birkhauser Boston.

\bibitem[{McDonald and Kaufman(1979)}]{MacDonold1979}
McDonald, S.~W., Kaufman, A.~N., 1979. Spectrum and eigenfunctions for a
  \textsc{H}amiltonian with stochastic trajectories. Phys. Rev. Lett. 42, 1189.

\bibitem[{Mehta(1991)}]{Mehta1991B}
Mehta, M.~L., 1991. Random Matrices, 2nd Edition. Academic Press, Boston.

\bibitem[{Mello et~al.(1985)Mello, Peveyra, and Seligman}]{Mello1985}
Mello, P.~A., Peveyra, P., Seligman, T.~H., 1985. Information theory and
  statistical nuclear reactions. \textsc{I}. general theory and applications to
  few-channel problems. Ann. of Phys. 161, 254.

\bibitem[{Mitchell et~al.(2010)Mitchell, Richter, and
  Weidenmuller}]{Mitchell01RMP_2010}
Mitchell, G.~E., Richter, A., Weidenmuller, H.~A., 2010. Rev.\ Mod.\ Phys. 82,
  2845.

\bibitem[{Moglie and Primiani(2012)}]{2012_Moglie}
Moglie, F., Primiani, V., april 2012. Numerical analysis of a new location for
  the working volume inside a reverberation chamber. Electromagnetic
  Compatibility, IEEE Transactions on 54~(2), 238 --245.

\bibitem[{Orjubin et~al.(2006)Orjubin, Richalot, Mengue, and
  Picon}]{2006_Picon}
Orjubin, G., Richalot, E., Mengue, S., Picon, O., feb. 2006. Statistical model
  of an undermoded reverberation chamber. Electromagnetic Compatibility, IEEE
  Transactions on 48~(1), 248 --251.

\bibitem[{Ott(2002)}]{Ott2002B}
Ott, E., 2002. Chaos in Dynamical Systems, 2nd Edition. Cambridge University
  Press, New York.

\bibitem[{Pirkl et~al.(2011)Pirkl, Ladbury, and Remley}]{2011_Ladbury}
Pirkl, R., Ladbury, J., Remley, K., aug. 2011. The reverberation chamber's
  unstirred field: A validation of the image theory interpretation. In:
  Electromagnetic Compatibility (EMC), 2011 IEEE International Symposium on.
  pp. 670 --675.

\bibitem[{Prange(2005)}]{Prange2005}
Prange, R.~E., 2005. Resurgence in quasi-classical scattering. J. Phys. A:
  Math. Gen. 38, 10703.

\bibitem[{Primiani et~al.(2009)Primiani, Moglie, and Paolella}]{2009_Moglie}
Primiani, V., Moglie, F., Paolella, V., aug. 2009. Numerical and experimental
  investigation of unstirred frequencies in reverberation chambers. In:
  Electromagnetic Compatibility, 2009. EMC 2009. IEEE International Symposium
  on. pp. 177 --181.

\bibitem[{Primiani and Moglie(2010)}]{2010_Mariani}
Primiani, V.~M., Moglie, F., 2010. Numerical simulation of los and nlos
  conditions for an antenna inside a reverberation chamber. Journal of
  Electromagnetic Waves and Applications 24~(17-18), 2319--2331.

\bibitem[{Savin et~al.(2001)Savin, Fyodorov, and Sommers}]{Savin2001}
Savin, D.~V., Fyodorov, Y.~V., Sommers, H.-J., 2001. Reducing nonideal to ideal
  coupling in random matrix description of chaotic scattering: Application to
  the time-delay problem. Phys. Rev. E 63, 035202(R).

\bibitem[{Savin and Sommers(2004)}]{Savin2004}
Savin, D.~V., Sommers, H.-J., 2004. Distribution of reflection eigenvalues in
  many-channel chaotic cavities with absorption. Phys. Rev. E 69, 035201(R).

\bibitem[{Savin et~al.(2005)Savin, Sommers, and Fyodorov}]{Savin2005}
Savin, D.~V., Sommers, H.-J., Fyodorov, Y.~V., 2005. Universal statistics of
  the local \textsc{G}reen's function in wave chaotic systems with absorption.
  JETP Lett. 82, 544.

\bibitem[{Sch\"afer et~al.(2005)Sch\"afer, St\"ockmann, Gorin, and
  Seligman}]{Schafer2005_2}
Sch\"afer, R., St\"ockmann, H.-J., Gorin, T., Seligman, T.~H., Oct 2005.
  Experimental verification of fidelity decay: From perturbative to fermi
  golden rule regime. Phys. Rev. Lett. 95, 184102.

\bibitem[{Schomerus and Jacquod(2005)}]{Jacquod2005}
Schomerus, H., Jacquod, P., 2005. Quantum-to-classical correspondence in open
  chaotic systems. Journal of Physics A: Mathematical and General 38~(49),
  10663.

\bibitem[{Schwabl and Brewer(2006)}]{schwabl2006}
Schwabl, F., Brewer, W., 2006. Statistical Mechanics. Advanced Texts in
  Physics. Springer.
\newline\urlprefix\url{http://books.google.com/books?id=7VnKAW284PgC}

\bibitem[{So et~al.(1995)So, Anlage, Ott, and Oerter}]{So_1995}
So, P., Anlage, S.~M., Ott, E., Oerter, R.~N., Apr 1995. Wave chaos experiments
  with and without time reversal symmetry: Gue and goe statistics. Phys. Rev.
  Lett. 74, 2662--2665.

\bibitem[{Sridhar(1991)}]{Sridhar1991}
Sridhar, S., Aug 1991. Experimental observation of scarred eigenfunctions of
  chaotic microwave cavities. Phys. Rev. Lett. 67, 785--788.

\bibitem[{Stein and St\"{o}ckmann(1992)}]{Stein1992}
Stein, J., St\"{o}ckmann, H.-J., 1992. Experimental determination of billiard
  wave functions. Phys. Rev. Lett. 68, 2867.

\bibitem[{St\"{o}ckmann and Stein(1990)}]{Stockmann1990}
St\"{o}ckmann, H.-J., Stein, J., 1990. ``\textsc{Q}uantum'' chaos in billiards
  studied by microwave absorption. Phys. Rev. Lett. 64, 2215.

\bibitem[{Tanner et~al.(2010)Tanner, Chappell, Hamdin, Giani, Seidel, and
  Vogel}]{2010_Tanner}
Tanner, G., Chappell, D., Hamdin, H.~B., Giani, S., Seidel, C., Vogel, F.,
  2010. Acoustic energy distribution in multi-component structures - dynamical
  energy analysis versus numerically exact results. In: Proceedings of the ISMA
  2010 and USD 2010. pp. 2435 -- 2436.

\bibitem[{Tanner and Sondergaard(2007)}]{2007_Tanner}
Tanner, G., Sondergaard, N., 2007. Wave chaos in acoustics and elasticity. J.
  Phys. A 40~(50).

\bibitem[{T\"ureci et~al.(2006)T\"ureci, Stone, and Collier}]{2006Stone}
T\"ureci, H.~E., Stone, A.~D., Collier, B., Oct 2006. Self-consistent multimode
  lasing theory for complex or random lasing media. Phys. Rev. A 74, 043822.
\newline\urlprefix\url{http://link.aps.org/doi/10.1103/PhysRevA.74.043822}

\bibitem[{Urbina and Richter(2006)}]{Urbina2006}
Urbina, J.~D., Richter, K., 2006. Statistical description of eigenfunctions in
  chaotic and weakly disordered systems beyond universality. Phys. Rev. Lett.
  97, 214101.

\bibitem[{Warne et~al.(2003)Warne, Lee, Hudson, Johnson, Jorgenson, and
  Stronach}]{2003_Warne}
Warne, L., Lee, K., Hudson, H., Johnson, W., Jorgenson, R., Stronach, S., may
  2003. Statistical properties of linear antenna impedance in an electrically
  large cavity. Antennas and Propagation, IEEE Transactions on 51~(5), 978 --
  992.

\bibitem[{Weidenm\"uller and Mitchell(2009)}]{Weiden_RMP_2009}
Weidenm\"uller, H.~A., Mitchell, G.~E., May 2009. Random matrices and chaos in
  nuclear physics: Nuclear structure. Rev. Mod. Phys. 81, 539--589.

\bibitem[{Wigner(1967)}]{1967_Wigner}
Wigner, E., 1967. Random matrices in physics. SIAM Review 9~(1), 1--23.

\bibitem[{Wigner and Dirac(1950)}]{Wigner_Dirac_1950}
Wigner, E.~P., Dirac, P. A.~M., 1950. On the statistical distribution of the
  widths and spacings of nuclear resonance levels. Proceedings of the Cambridge
  Philosophical Society 47, 790.

\bibitem[{Wright and Weaver(2010)}]{Weaver_2010}
Wright, M., Weaver, R., 2010. New Directions in Linear Acoustics and Vibration:
  Quantum Chaos, Random Matrix Theory and Complexity. Cambridge University
  Press.

\bibitem[{Yeh et~al.(2012{\natexlab{a}})Yeh, Antonsen, Ott, and
  Anlage}]{Yeh2012b}
Yeh, J.-H., Antonsen, T.~M., Ott, E., Anlage, S.~M., 2012{\natexlab{a}}. Fading
  statistics in communications - a random matrix approach. Acta Physica
  Polonica A 120, A--85.

\bibitem[{Yeh et~al.(2012{\natexlab{b}})Yeh, Antonsen, Ott, and
  Anlage}]{Yeh2012a}
Yeh, J.-H., Antonsen, T.~M., Ott, E., Anlage, S.~M., 2012{\natexlab{b}}.
  First-principles model of time-dependent variations in transmission through a
  fluctuating scattering environment. Phys. Rev. E(Rapid Communications) 85,
  015202.

\bibitem[{Yeh et~al.(2010{\natexlab{a}})Yeh, Hart, Bradshaw, Antonsen, Ott, and
  Anlage}]{Yeh2010b}
Yeh, J.-H., Hart, J.~A., Bradshaw, E., Antonsen, T.~M., Ott, E., Anlage, S.~M.,
  2010{\natexlab{a}}. Experimental examination of the effect of short ray
  trajectories in two-port wave-chaotic scattering systems. Phys. Rev. E 82,
  041114.

\bibitem[{Yeh et~al.(2010{\natexlab{b}})Yeh, Hart, Bradshaw, Antonsen, Ott, and
  Anlage}]{Yeh2010a}
Yeh, J.-H., Hart, J.~A., Bradshaw, E., Antonsen, T.~M., Ott, E., Anlage, S.~M.,
  2010{\natexlab{b}}. Universal and nonuniversal properties of wave-chaotic
  scattering systems. Phys. Rev. E 81, 025201(R).

\bibitem[{Zheng(2005)}]{Zheng2005T}
Zheng, X., 2005. Statistics of impedance and scattering matrices in microwave
  chaotic cavities: the random coupling model. Ph.D. thesis, University of
  Maryland, College Park.

\bibitem[{Zheng et~al.(2006{\natexlab{a}})Zheng, Antonsen, and
  Ott}]{Zheng2006a}
Zheng, X., Antonsen, T.~M., Ott, E., 2006{\natexlab{a}}. Statistics of
  impedance and scattering matrices in chaotic microwave cavities: Single
  channel case. Electromagnetics 26, 3.

\bibitem[{Zheng et~al.(2006{\natexlab{b}})Zheng, Antonsen, and
  Ott}]{Zheng2006b}
Zheng, X., Antonsen, T.~M., Ott, E., 2006{\natexlab{b}}. Statistics of
  impedance and scattering matrices of chaotic microwave cavities with multiple
  ports. Electromagnetics 26, 37.

\bibitem[{Zheng et~al.(2006{\natexlab{c}})Zheng, Hemmady, Antonsen, Anlage, and
  Ott}]{Zheng2006c}
Zheng, X., Hemmady, S., Antonsen, T.~M., Anlage, S.~M., Ott, E.,
  2006{\natexlab{c}}. Characterization of fluctuations of impedance and
  scattering matrices in wave chaotic scattering. Phys. Rev. E 73, 046208.

\end{thebibliography}

%% Authors are advised to submit their bibtex database files. They are
%% requested to list a bibtex style file in the manuscript if they do
%% not want to use elsarticle-harv.bst.

%% References without bibTeX database:

% \begin{thebibliography}{00}

%% \bibitem must have one of the following forms:
%%   \bibitem[Jones et al.(1990)]{key}...
%%   \bibitem[Jones et al.(1990)Jones, Baker, and Williams]{key}...
%%   \bibitem[Jones et al., 1990]{key}...
%%   \bibitem[\protect\citeauthoryear{Jones, Baker, and Williams}{Jones
%%       et al.}{1990}]{key}...
%%   \bibitem[\protect\citeauthoryear{Jones et al.}{1990}]{key}...
%%   \bibitem[\protect\astroncite{Jones et al.}{1990}]{key}...
%%   \bibitem[\protect\citename{Jones et al., }1990]{key}...
%%   \harvarditem[Jones et al.]{Jones, Baker, and Williams}{1990}{key}...
%%

% \bibitem[ ()]{}

% \end{thebibliography}
\end{document}